\documentclass[aps,twocolumn,superscriptaddress,nofootinbib,amsmath,amssymb,floatfix]{revtex4-2}

\usepackage{mathtools}
\usepackage{bm}
\usepackage{graphicx}
\usepackage{hyperref}
\usepackage{xcolor}

\graphicspath{{./figures/}}

\newcommand{\E}{\mathbb{E}}
\newcommand{\Var}{\mathrm{Var}}
\newcommand{\Cov}{\mathrm{Cov}}

\newcommand{\diag}{\mathrm{diag}}
\newcommand{\sech}{\mathrm{sech}}
\newcommand{\He}{\mathrm{He}}
\newcommand{\dd}{\mathrm{d}}
\newcommand{\re}{\mathrm{Re}}
\newcommand{\im}{\mathrm{Im}}

\begin{document}

\title{Mean-field theory of rich oscillatory dynamics in low-rank recurrent networks with activity-dependent adaptation}
\author{Bowen Zheng}
\affiliation{Department of Brain and Cognitive Sciences, Massachusetts Institute of Technology, Cambridge, MA 02139, USA}

\author{Earl K. Miller}
\thanks{E.K.M.\ and I.R.F.\ jointly supervised this work.}
\affiliation{Department of Brain and Cognitive Sciences, Massachusetts Institute of Technology, Cambridge, MA 02139, USA}
\affiliation{The Picower Institute for Learning and Memory, Massachusetts Institute of Technology, Cambridge, MA 02139, USA}

\author{Ila R. Fiete}
\email{fiete@mit.edu}
\affiliation{Department of Brain and Cognitive Sciences, Massachusetts Institute of Technology, Cambridge, MA 02139, USA}
\affiliation{McGovern Institute for Brain Research, Massachusetts Institute of Technology, Cambridge, MA 02139, USA}
\date{\today}

\begin{abstract}
We develop a dynamical mean-field theory for random recurrent networks with
low-rank structure and firing-rate-driven adaptation. When the random
connectivity is strong enough to generate chaos, increasing adaptation strength
drives the network through four regimes: a static coherent state,
noise-sustained oscillations that progress from regular to irregular,
stochastic switching between symmetric wells, and a global limit cycle. The
theory identifies two instability mechanisms, chaos onset from the random
connectivity and a Hopf bifurcation of the coherent mode, and shows how
adaptation shapes both through the frequency-dependent single-neuron transfer
function. A reduced three-dimensional model
captures the bifurcation structure of the full network. Above the chaos
threshold, coherent population-level oscillations coexist with heterogeneous
firing rates and network-generated stochasticity at the single-neuron level. The interaction of adaptation with random and low-rank connectivity produces
a rich oscillatory repertoire, including waxing-and-waning rhythmic episodes,
persistent state switching, and slow Up-Down alternations, dynamics that
have been observed during wakefulness, sleep, and anesthesia.
\end{abstract}

\maketitle

\section{Introduction}
\label{sec:intro}

Oscillatory activity is ubiquitous in the brain. Rhythms spanning frequencies from slow oscillations during sleep to gamma-band activity during sensory processing organize neural computation, gate information flow between areas, and coordinate the timing of spikes across populations~\cite{Buzsaki2004,Fries2005,WangBuzsaki1996}. An interesting feature of these rhythms is their coexistence with irregular, heterogeneous firing at the single-neuron level~\cite{Wang2010,Buzsaki2004}. Diverse circuit mechanisms have been proposed to account for specific oscillatory phenomena: recurrent excitatory-inhibitory feedback for gamma rhythms~\cite{WilsonCowan1972,WangBuzsaki1996,BrunelWang2003}, thalamocortical loops for sleep spindles~\cite{Destexhe1993}, and slow adaptation currents for the cortical slow oscillation~\cite{Compte2003}. Each successfully explains a particular rhythm but requires specific cell types, synaptic kinetics, and connectivity motifs. A complementary tradition studies large random networks of rate neurons, which naturally produce irregular and heterogeneous firing through a transition to deterministic chaos~\cite{SCS88,vanVreeswijkSompolinsky1996,CrisantiSompolinsky2018}. These networks capture single-neuron irregularity but generate no oscillations and no spatial structure in their fluctuations. In sparsely connected spiking networks, Brunel~\cite{Brunel2000} showed that excitatory-inhibitory balance can produce both population-wide oscillations and single-neuron irregularity, mapping a phase diagram with multiple dynamical regimes including synchronous irregular firing. Whether a comparably rich oscillatory phase structure can arise within a minimal rate-network framework remains open.

Within the rate-network framework, low-rank perturbations of random connectivity have emerged as a natural model class for studying how low-dimensional dynamics arise from recurrent circuits~\cite{MO18,Kadmon2015,Schuessler2020,LandauSompolinsky2018,LandauSompolinsky2021,AljadeffSternSharpee2015}, consistent with the low-dimensional structure widely observed in population recordings~\cite{CunninghamYu2014,Gallego2017} and supporting a growing theoretical framework for low-dimensional computation~\cite{Beiran2021,Dubreuil2022,SussilloAbbott2009,Ostojic2014,RivkindBarak2017,ClarkAbbottLitwinKumar2023}. Oscillations can arise in these networks, but through connectivity structure: in rank-one networks with orthogonal input-output modes, Landau and Sompolinsky~\cite{LandauSompolinsky2018} showed that coherent oscillatory chaos occurs for connectivity realizations whose leading eigenvalue is complex, with the oscillation frequency determined by the random bulk. Mastrogiuseppe and Ostojic~\cite{MO18} showed that rank-one networks produce only fixed-point dynamics, and that oscillations require rank-two connectivity with appropriate geometry. In both cases, whether and at what frequency the network oscillates is controlled by the connectivity structure rather than by a single-neuron property. An alternative is to endow individual neurons with a slow intrinsic timescale. Spike-frequency adaptation (SFA), a progressive reduction in firing rate under sustained input mediated by slow potassium currents~\cite{BendaHerz2003,Fuhrmann2002,Ermentrout2001,vanVreeswijkHansel2001}, is a natural candidate. Muscinelli, Gerstner, and Schwalger~\cite{MGS19} showed that adaptation in purely random networks introduces a narrow-band spectral peak at the adaptation frequency, termed ``resonant chaos,'' but these fluctuations remain spatially unstructured with no population-wide organization. Clark and Abbott~\cite{ClarkAbbott2024} showed that slow variables interacting with fast neuronal dynamics in random networks can produce rich phase diagrams, but again without low-dimensional population structure. Neither low-rank connectivity alone nor adaptation alone produces population-wide oscillations across multiple dynamical regimes.

The present work combines low-rank connectivity with firing-rate-driven adaptation in a random rate network. We show that this combination is sufficient to produce a rich oscillatory repertoire, organized by the interplay of adaptation strength, structural coupling, and chaos intensity, while preserving single-neuron irregularity from the chaotic background. We adopt the adaptation equation
\begin{equation}
\tau_a \dot{a} = -a + \beta\,\phi(x), \label{eq:intro_adapt}
\end{equation}
where $\phi(x) = \tanh(x)$ is the firing rate, reflecting the dominant biophysical mechanisms of SFA in which spike-triggered potassium currents provide negative feedback on the firing rate~\cite{BendaHerz2003, GollischHerz2004}. This form also gives $\beta$ a natural interpretation as an
effective adaptation strength that can be modulated by neuromodulatory state.
Cholinergic modulation can suppress slow potassium currents associated with
spike-frequency adaptation~\cite{mccormick1993actions,stiefel2008cholinergic}. Conversely,
reduced cholinergic drive has been linked to stronger effective adaptation and
distinct slow-wave dynamics in sleep and anesthesia~\cite{nghiem2020cholinergic}. This
motivates asking whether changes in a single adaptation parameter can organize
transitions among irregular wake-like activity, intermittent rhythmic episodes,
and slow Up--Down alternations.

The low-rank structure creates a coherent mode in which each neuron fires at a heterogeneous rate set by its projection onto the connectivity pattern. Increasing adaptation strength drives the network from this static coherent state into local oscillations around these fixed-point rates. In this regime the coherent mode is a stable focus, functioning as a damped oscillator at a frequency set by the adaptation timescale. When the random connectivity also generates a chaotic background, this self-generated noise selectively drives the coherent mode at its resonant frequency, producing sustained population-level oscillations even though the coherent mode remains linearly stable. We term this noise-sustained coherent oscillation. With stronger adaptation the network can exhibit irregular switching between symmetric fixed points, driven by chaotic fluctuations accumulating through the slow adaptation variable, before eventually entering global oscillations with Up and Down states. The theory identifies two instability mechanisms: a chaotic transition driven by the random connectivity, and a Hopf bifurcation of the coherent mode driven by adaptation. Which instability occurs first determines the route through the phase diagram. If the coherent mode loses stability before chaos onset, the network jumps directly to global oscillations. If chaos occurs first, the network passes through a structured sequence of noise-sustained oscillatory regimes before reaching global oscillation. A reduced three-dimensional model captures the bifurcation structure and explains the progression across regimes through a state-dependent noise mechanism.

\section{Model}
\label{sec:model}

We consider $N$ rate neurons with membrane potential $x_i$ and adaptation current $a_i$. The dynamics are
\begin{align}
\tau_m\,\dot{x}_i &= -x_i + \sum_{j=1}^N J_{ij}\,\phi(x_j) - a_i, \label{eq:model_x}\\
\tau_a\,\dot{a}_i &= -a_i + \beta\,\phi(x_i), \label{eq:model_a}
\end{align}
with $\phi(x) = \tanh(x)$ and $\beta \geq 0$. We set $\tau_m = 1$ throughout and work in the regime $\tau_a \gg 1$.

We assume the connectivity decomposes into a random background and a rank-one structured component,
\begin{equation}
J_{ij} = \frac{g}{\sqrt{N}}\,W_{ij} + \frac{m_i\,n_j}{N},
\label{eq:connectivity}
\end{equation}
where $W_{ij} \overset{\mathrm{iid}}{\sim} \mathcal{N}(0,1)$ and the loading vectors are drawn jointly as $(m_i, n_i) \overset{\mathrm{iid}}{\sim}\mathcal{N}(0,\Sigma_{mn})$ with
\begin{equation}
\Sigma_{mn} = \begin{pmatrix} \sigma_m^2 & \gamma\sigma_m\sigma_n \\ \gamma\sigma_m\sigma_n & \sigma_n^2 \end{pmatrix},
\end{equation}
independent of $W$. Equivalently, $m_i = \sigma_m u_i$ and $n_i = \sigma_n(\gamma u_i + \sqrt{1-\gamma^2}\,\zeta_i)$ with $u_i,\zeta_i\overset{\mathrm{iid}}{\sim}\mathcal{N}(0,1)$. The effective outlier eigenvalue is $\lambda_{\mathrm{eff}} = \gamma\sigma_m\sigma_n$ \cite{LandauSompolinsky2018, MO18}. Unless stated otherwise, all simulations use $\sigma_m = \sigma_n = 2.0$, $\gamma = 0.7$, $\tau_a = 30$, and $N = 4000$.

The rank-one structure induces a coherent population mode described by the overlap
\begin{equation}
\kappa(t) = \frac{1}{N}\sum_{j=1}^N n_j\,\phi(x_j(t)),
\end{equation}
and we define the corresponding adaptation overlap 
\begin{equation}
\kappa_a(t) = N^{-1}\sum_j n_j\,a_j(t),
\end{equation} which from~\eqref{eq:model_a} satisfies the exact equation $\tau_a\,\dot{\kappa}_a = -\kappa_a + \beta\,\kappa$. The incoherent fluctuations are characterized by the mean-squared firing rate $Q(t) = N^{-1}\sum_j \phi(x_j(t))^2$. The pair $(\kappa, \kappa_a)$ describes the coherent mode and $Q$ describes the chaotic background; the interplay between them is the main subject of this paper. 

We focus on rank-one structure for conceptual clarity. Rank one is the minimal low-rank perturbation that creates a coherent population
mode. However, a single mode carries no rotational geometry and cannot, on its own, sustain a closed orbit. Higher-rank connectivity
can build oscillations in directly. For instance, in a rank-two network, the overlaps among the two pairs of loading vectors define an effective $2\times2$ coupling on the latent variables, and a rotational component of this
coupling would endow the structured connectivity with a complex-conjugate pair of
outlier eigenvalues, producing a limit cycle whose frequency is set by the
connectivity itself~\cite{MO18}. Oscillations of this kind are by construction a
property of the connectivity geometry. By contrast, the regimes studied here have a different mechanism for oscillatory dynamics. With a single coherent mode and no rotational structure in $J$, the oscillations are not designed into the connectivity but arise from the interaction of that mode with single-neuron adaptation and the chaotic random bulk. Higher-rank structure could layer connectivity-driven oscillatory mechanisms on top of the adaptation-driven ones described here, as discussed in Sec.~\ref{sec:extensions}.

\section{Dynamical regimes}
\label{sec:pheno}

\begin{figure*}[t]
\centering
\includegraphics[width=\textwidth]{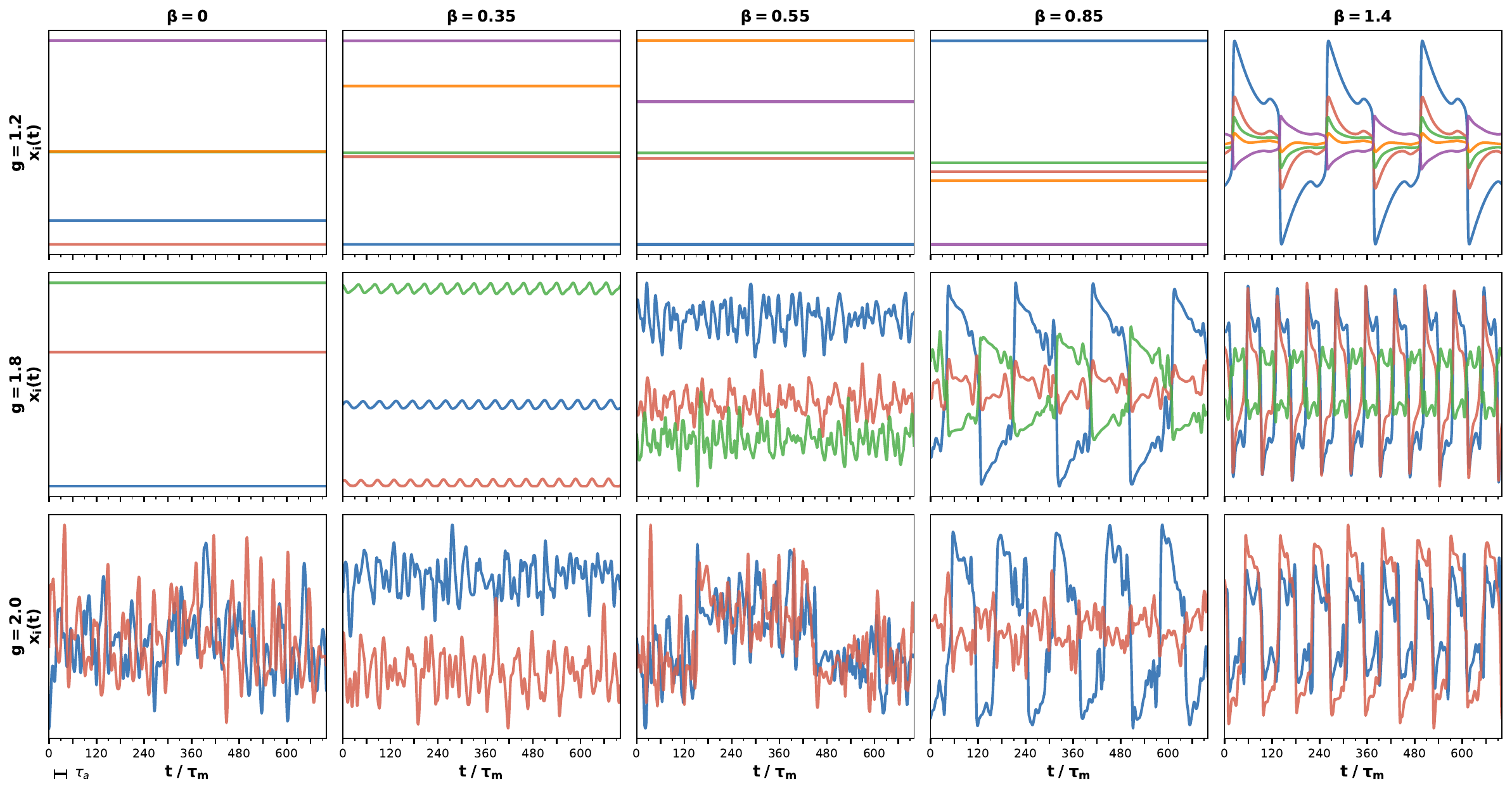}
\caption{\textbf{Adaptation strength and random connectivity organize distinct dynamical regimes.}
The panels show individual neuron time traces $x_i(t)$ across dynamical regimes.
Rows correspond to increasing random coupling strength $g = 1.2$ (below chaos),
$1.8$, and $2.0$ (both above chaos). Columns correspond to increasing
adaptation strength $\beta = 0, 0.35, 0.55, 0.85, 1.4$.
At $g=1.2$, the coherent mode loses stability before chaos onset: the network
passes directly from static fixed points to global oscillation (Regime~IV) at
large $\beta$. At $g=1.8$ and $g=2.0$, the network is chaotic and the full
sequence of regimes is visible: noise-sustained oscillation (Regime~II),
irregular switching between wells (Regime~III), and global oscillation
(Regime~IV). Time is shown in units of $\tau_m$, with $\tau_a=30$.
Parameters: $N = 4000$, $\tau_a = 30$, $\sigma_m = \sigma_n = 2.0$,
$\gamma = 0.7$.}
\label{fig:neuron_activity}
\end{figure*}

We describe the dynamical regimes observed in network simulations. The progression depends on the order in which two instabilities occur as $\beta$ increases: chaos from the random connectivity, and a Hopf bifurcation of the coherent mode from adaptation. We first describe the regime sequence for $g$ above the chaos threshold (Fig.~\ref{fig:neuron_activity}, second and third rows), where the random connectivity generates a chaotic background before the coherent mode loses stability, and then note the alternative route at lower $g$.

\begin{figure*}[t]
\centering
\includegraphics[width=\textwidth]{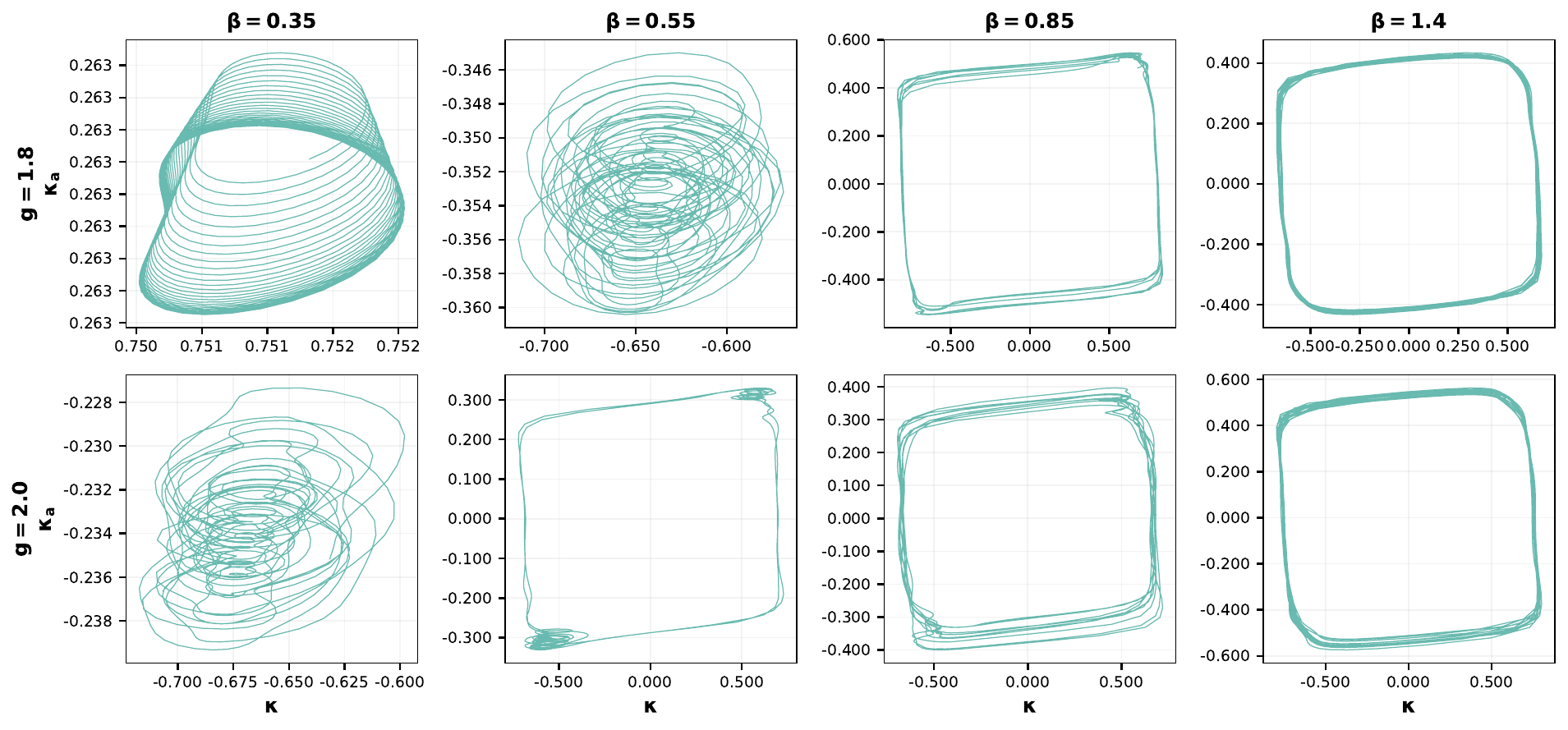}
\caption{\textbf{The coherent mode progresses from local oscillations to
switching and global cycles.}
The panels show phase portraits of the coherent overlap $\kappa(t)$ and the
adaptation overlap $\kappa_a(t)$ from network simulations. Top row: $g = 1.8$; bottom row: $g = 2.0$. Columns from left to
right: $\beta = 0.35$, $0.55$, $0.85$, $1.4$.
At $g = 1.8$: regular then irregular noise-sustained oscillation within one
basin; global oscillation with lobes around the unstable fixed points; smooth
limit cycle.
At $g = 2.0$, stronger chaos advances the sequence: irregular oscillation;
switching between basins; limit cycle with dwelling near the basin;
smooth relaxation oscillation.
Parameters as in Fig.~\ref{fig:neuron_activity}.}
\label{fig:order_params_2d}
\end{figure*}

For $g$ above the chaos threshold, increasing $\beta$ reveals four successive regimes (Figs.~\ref{fig:neuron_activity} and~\ref{fig:order_params_2d}).

\textit{Regime I: Static coherent state.}
The overlap $\kappa(t)$ settles near one of two symmetric fixed points $\pm\kappa^*$ with small fluctuations. The fixed points are stable nodes.

\textit{Regime II: Noise-sustained oscillation.}
The coherent fixed points become stable foci. The chaotic background acts as a
broadband driving force that sustains oscillations of $\kappa(t)$ around one of
the symmetric fixed points. This oscillatory fluctuation is absent when $g$ is
reduced below the chaos threshold. At moderate adaptation strength
(e.g. $g = 1.8, \beta=0.35$ in Fig.~\ref{fig:order_params_2d}), the oscillations are
approximately periodic. As $\beta$ increases further
(e.g. $g = 1.8, \beta=0.55$), the oscillations become irregular.

\textit{Regime III: Irregular switching.}
As $\beta$ increases, the coherent fixed points approach a stability boundary.
Chaotic fluctuations, accumulating through the slow adaptation variable
$\kappa_a$, occasionally drive the trajectory from one well to the other,
producing irregular transitions between the two symmetric states
(e.g. $g = 2.0, \beta=0.85$ in Fig.~\ref{fig:order_params_2d}). The frequency of
switching increases with $\beta$.

\textit{Regime IV: Global oscillation.}
For large $\beta$ (e.g. $\beta=1.4$ in
Fig.~\ref{fig:order_params_2d}), the coherent mode undergoes a Hopf
bifurcation and a stable limit cycle appears, carrying $\kappa(t)$
periodically between the two wells. Above the chaos threshold the limit cycle
coexists with chaotic fluctuations in the individual neurons.

The mechanism underlying Regime~II can be understood by analogy with a damped oscillator driven by broadband noise. The oscillator produces sustained fluctuations concentrated near its resonant frequency, even though it is linearly stable. Here the coherent mode plays the role of the oscillator, with adaptation providing the restoring force, and the chaotic bulk plays the role of the noise source. Neither alone produces spatially coherent oscillations. Without chaos, the coherent mode sits at a stable fixed point after initial damping oscillation; without the low-rank structure, the chaos is spatially unstructured and produces no organized population-level oscillation. The quantitative theory is developed in Secs.~\ref{sec:spectral}--\ref{sec:coherent}.

When $g$ is below the chaos threshold, the alternative route is observed (Fig.~\ref{fig:neuron_activity}, top row). As $\beta$ increases, the coherent mode loses stability through a Hopf bifurcation before chaos sets in. The network passes directly from the static coherent state (Regime~I) to global oscillation (Regime~IV), without the intermediate noise-sustained regimes, since there is no chaotic background to drive them.

The remainder of the paper develops a mean-field theory that accounts for these regimes and the two routes through the phase diagram. We first derive the DMFT fixed-point statistics, which determine how adaptation reshapes the distribution of neuronal operating points (Sec.~\ref{sec:dmft}). We then analyze the chaos boundary set by the bulk eigenvalues of the random connectivity (Sec.~\ref{sec:spectral}) and compute the self-consistent fluctuation spectrum above this boundary (Sec.~\ref{sec:spectrum}). We then analyze the stability of the coherent mode in the presence of bulk noise, which determines the Hopf boundary and completes the phase diagram in the $(\beta, g)$ plane (Sec.~\ref{sec:coherent}). Finally, we construct a reduced three-dimensional dynamical model that captures the bifurcation structure and the noise-sustained oscillation mechanism (Sec.~\ref{sec:reduced}), and outline extensions to general rank and multiple adaptation channels (Sec.~\ref{sec:extensions}).

\section{Mean-field theory and fixed-point statistics}
\label{sec:dmft}

\subsection{DMFT equations}

This subsection derives the effective single-neuron process and the
self-consistency equations used for the fixed-point and fluctuation analyses.

Substituting~\eqref{eq:connectivity} into~\eqref{eq:model_x}, the total synaptic input to neuron $i$ separates as
\begin{equation}
\sum_j J_{ij}\,\phi(x_j) = m_i\,\kappa(t) + g\,\eta_i(t),
\label{eq:input_decomp}
\end{equation}
where $\eta_i(t) = N^{-1/2}\sum_j W_{ij}\phi(x_j(t))$. In the large-$N$ limit, $\eta_i$ is described by a Gaussian process, independent
across neurons to leading order, with zero mean and autocovariance
\begin{equation}
\begin{split}
&\E[\eta_i(t)\,\eta_i(t')] \xrightarrow{N\to\infty} R_\phi(t,t')\\
&\quad\equiv \frac{1}{N}\sum_{j=1}^N\phi(x_j(t))\,\phi(x_j(t')).
\end{split}
\label{eq:eta_cov}
\end{equation}
In the stationary regime $R_\phi(t,t') = R_\phi(t-t')$, and the equal-time value is $Q = R_\phi(0)$.

Each neuron in the $N\to\infty$ limit evolves as
\begin{align}
\tau_m\,\dot{x}_i &= -x_i - a_i + m_i\,\kappa(t) + g\,\eta_i(t), \label{eq:dmft_x}\\
\tau_a\,\dot{a}_i &= -a_i + \beta\,\phi(x_i), \label{eq:dmft_a}
\end{align}
with the self-consistency
\begin{equation}
R_\phi(\tau) = \E_h\!\big[\langle\phi_i(t)\,\phi_i(t+\tau)\rangle_t\big],
\label{eq:selfcons}
\end{equation}
where $\langle\cdot\rangle_t$ denotes a temporal average and $\E_h$ averages over the total stationary input $h_i = m_i\kappa + g\eta_i^{(0)}$, with $\eta_i^{(0)} \equiv N^{-1/2}\sum_j W_{ij}\langle\phi_j\rangle_t$ the time-averaged contribution from the random connectivity (the superscript $(0)$ distinguishes this DC component from the full process $\eta_i(t)$). The centered autocovariance is $C_\phi(\tau) = R_\phi(\tau) - \E_h[\langle\phi_i\rangle_t^2]$, with Fourier transform denoted by $S_\phi$. Across the population, $h_i \sim \mathcal{N}(0,\Sigma_h^2)$, where $\Sigma_h^2$ denotes the total input variance. At the stationary fixed point, where $\langle\phi_i\rangle_t = \phi(x_i^*)$ and $\Var(\eta_i^{(0)}) = Q$, it decomposes exactly into coherent and incoherent contributions:
\begin{equation}
\Sigma_h^2 = \sigma_m^2\,\kappa^2 + g^2\,Q.
\label{eq:Sigma_h}
\end{equation}
Above the chaos threshold, $Q = R_\phi(0)$ includes temporal fluctuations while $h_i$ is built from the time-averaged component $\eta_i^{(0)}$; Eq.~\eqref{eq:Sigma_h} then serves as a self-consistent fixed-point approximation.

\subsection{Fixed-point diffeomorphism and local gain}
We first analyze the stationary coherent state and show how adaptation reshapes the distribution of neuronal operating points.
At the stationary fixed point, $a_i^* = \beta\,\phi(x_i^*)$ and setting $\dot{x}_i = 0$ in~\eqref{eq:model_x} gives
\begin{equation}
\underbrace{x_i^* + \beta\,\tanh(x_i^*)}_{F(x_i^*)} = h_i.
\label{eq:fp}
\end{equation}
Since $F'(x) = 1 + \beta\sech^2(x) \geq 1$, the map $F:\mathbb{R}\to\mathbb{R}$ is a diffeomorphism and $x_i^* = F^{-1}(h_i)$ is unique. We define the \emph{local gain}
\begin{equation}
c_i \equiv \sech^2(x_i^*),
\label{eq:local_gain_def}
\end{equation}
which we also write as $c(h) = \sech^2(F^{-1}(h))$ when viewed as a function of the quenched input. The inverse derivative is
\begin{equation}
(F^{-1})'(h) = \frac{1}{1 + \beta\,c(h)}.
\label{eq:Finv_deriv}
\end{equation}

This diffeomorphism distinguishes our model from the adaptation-free case, where $F(x) = x$. It encodes how adaptation compresses the neuronal operating range (Fig.~\ref{fig:density_fp}). Neurons receiving large input $|h_i| \gg 1$ have $c_i \approx 0$, so $F^{-1}(h) \approx h$ and adaptation has little effect. Near the center of the population, $c_i$ is close to unity and adaptation maximally compresses the gain by a factor $1/(1+\beta)$.

Since $h \sim \mathcal{N}(0,\Sigma_h^2)$, population averages over the gain are one-dimensional Gaussian integrals. We define the population-averaged susceptibilities
\begin{align}
\bar{\chi}_x &= \E_h\!\big[c(h)\big], \label{eq:chi_x}\\
\bar{\chi}_{\mathrm{eff}} &= \E_h\!\left[\frac{c(h)}{1+\beta\,c(h)}\right], \label{eq:chi_eff_def}
\end{align}
and the population-averaged squared gain $\chi_{2,x} = \E_h[c(h)^2]$, which will set the effective coupling strength for chaos onset (Sec.~\ref{sec:spectral}).

\subsection{Overlap self-consistency}
\begin{figure*}[t]
\centering
\includegraphics[width=0.85\textwidth]{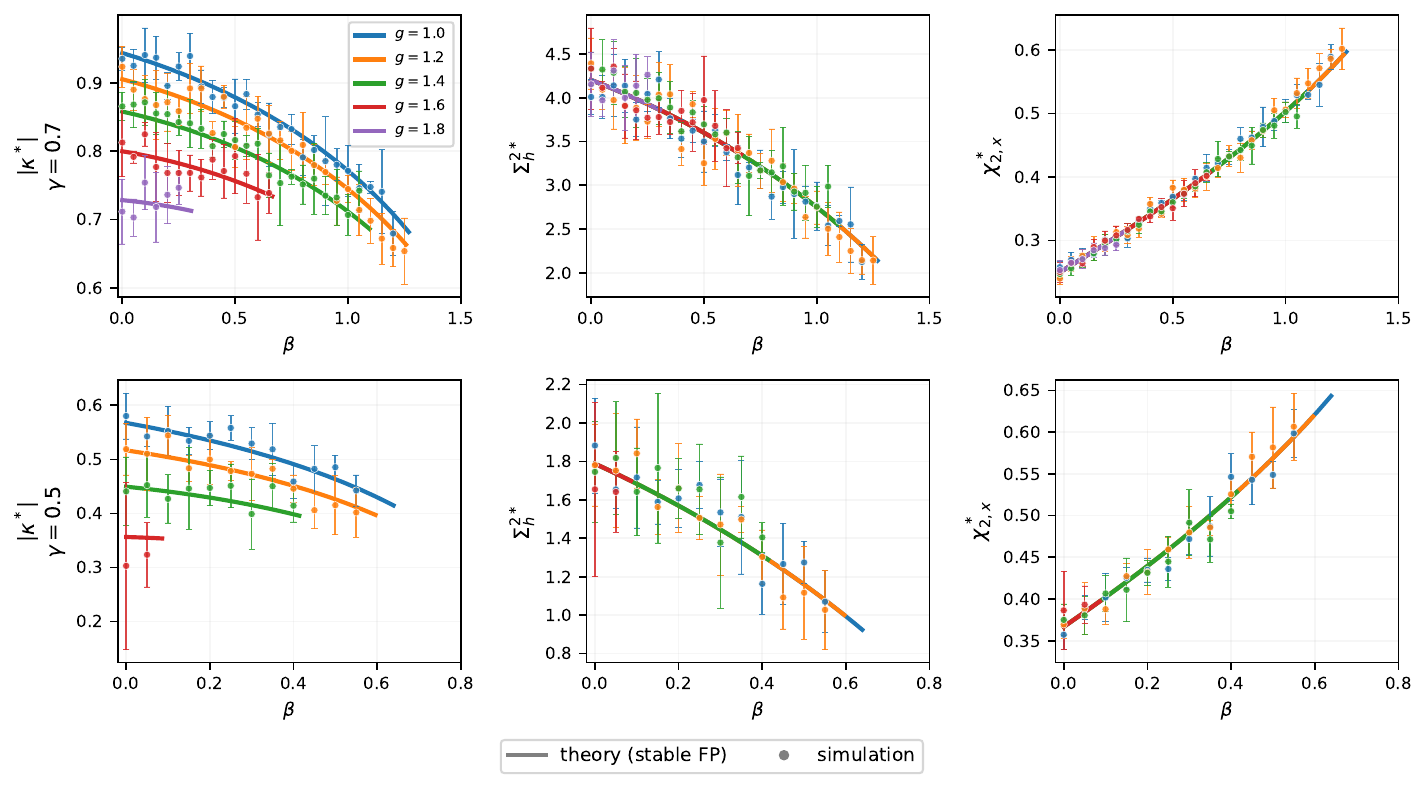}
\caption{\textbf{DMFT predicts the coherent fixed-point branch and gain
statistics.}
Top row: $\gamma = 0.7$; bottom row: $\gamma = 0.5$.
Left: overlap magnitude $|\kappa^*|$ versus $\beta$.
Center: input variance $\Sigma_h^{2*}$ versus $\beta$.
Right: fourth-moment susceptibility $\chi_{2,x}^*$ versus $\beta$.
Solid curves: theory (stable fixed-point branch).
Points with error bars: simulations at $N = 4000$, averaged over 5 random realizations.
Different colors correspond to $g = 1.0, 1.2, 1.4, 1.6, 1.8$.
The curves terminate where the fixed point ceases to exist.
Parameters: $\sigma_m = \sigma_n = 2.0$, $\tau_a = 30$.}
\label{fig:fp_validation}
\end{figure*}

We now close the fixed-point theory by deriving the self-consistency condition for the coherent overlap.
Across the population, $h_i \sim \mathcal{N}(0, \Sigma_h^2)$. The overlap $\kappa = N^{-1}\sum_j n_j\,\phi(x_j^*)$ is computed via Stein's lemma. Since $(n_j, h_j)$ are jointly Gaussian with $\Cov(n,h) = \lambda_{\mathrm{eff}}\kappa$, and $\phi(x^*) = \tanh(F^{-1}(h))$ is a function of $h$ alone, the lemma gives $\kappa = \lambda_{\mathrm{eff}}\kappa\,\E[\psi'(h)]$ where $\psi(h) \equiv \tanh(F^{-1}(h))$ is the fixed-point firing rate as a function of quenched input. The chain rule yields
\begin{equation}
\psi'(h) = \frac{c(h)}{1 + \beta\,c(h)},
\label{eq:psi_prime}
\end{equation}
so that $\kappa = \lambda_{\mathrm{eff}}\kappa\,\bar{\chi}_{\mathrm{eff}}(\Sigma_h^2,\beta)$. For $\kappa \neq 0$ the self-consistency condition is
\begin{equation}
\lambda_{\mathrm{eff}}\,\bar{\chi}_{\mathrm{eff}}(\Sigma_h^2, \beta) = 1.
\label{eq:selfcons_fp}
\end{equation}
Together with the variance condition $Q = Q_{\mathrm{fp}}(\Sigma_h^2,\beta) \equiv \E_Z[\tanh^2(F^{-1}(\Sigma_h Z))]$, this determines the fixed-point values of $\Sigma_h^2$, $\kappa$, and $Q$. The overlap follows from $\kappa^2 = (\Sigma_h^2 - g^2 Q)/\sigma_m^2$. Since $\bar{\chi}_{\mathrm{eff}}$ decreases monotonically with $\Sigma_h^2$ and equals $1/(1+\beta)$ at $\Sigma_h^2 = 0$, a nonzero overlap requires $\lambda_{\mathrm{eff}} > 1 + \beta$. Figure~\ref{fig:fp_validation} validates these fixed-point predictions against network simulations for two values of $\gamma$. Equation~\eqref{eq:selfcons_fp} determines the location of the nonzero coherent fixed-point branch. The linear stability of this branch is determined separately by the coherent loop gain derived in Sec.~\ref{sec:coherent}.

\begin{figure*}[t]
\centering
\includegraphics[width=\textwidth]{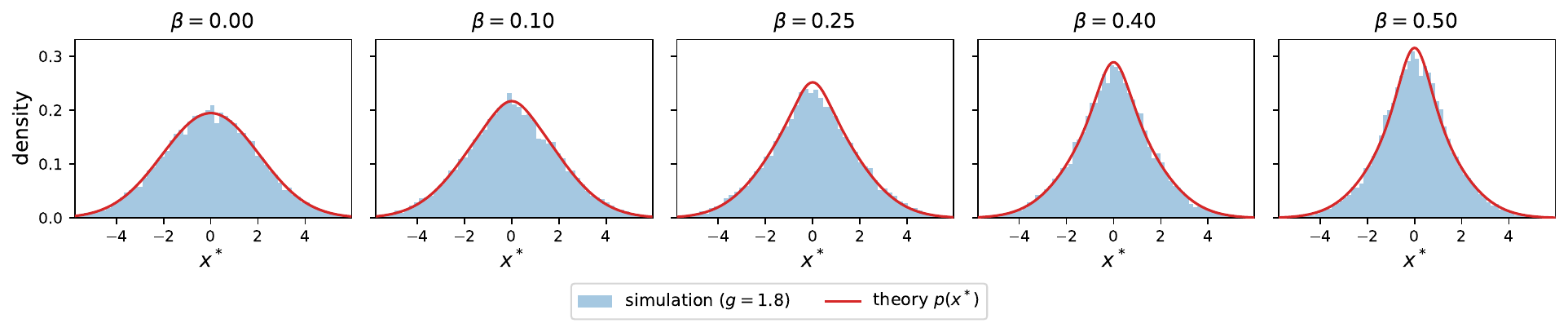}
\caption{\textbf{Adaptation concentrates fixed points near high gain and lowers
the chaos threshold.}
The panels show the fixed-point membrane-potential distribution $p(x^*)$.
Histograms show network simulations at $g=1.8$ and $N=4000$. Solid curves show
the DMFT prediction
$p(x^*) = p_h(F(x^*))\,F'(x^*)$, where
$p_h = \mathcal{N}(0,\Sigma_h^{2*})$. As the adaptation strength $\beta$
increases, the adaptation-induced diffeomorphism $F(x)=x+\beta\tanh(x)$
compresses the distribution toward zero. This compression concentrates neurons
in the high-gain region, increases the squared-gain susceptibility
$\chi_{2,x}$, and thereby lowers the chaos onset threshold
$g_c \propto 1/\sqrt{\chi_{2,x}}$. Parameters are as in
Fig.~\ref{fig:neuron_activity}.}
\label{fig:density_fp}
\end{figure*}

Figures~\ref{fig:density_fp} and~\ref{fig:fp_validation} together illustrate how adaptation facilitates chaos onset. As $\beta$ increases, the diffeomorphism $F$ compresses the fixed-point distribution $p(x^*)$ toward zero (Fig.~\ref{fig:density_fp}), concentrating neurons in the high-gain region where $c_i$ is large. This raises the population-averaged squared gain $\chi_{2,x}$ (Fig.~\ref{fig:fp_validation}, right panels). Since the chaos threshold scales as $g_c \propto 1/\sqrt{\chi_{2,x}}$ (derived in the next section), stronger adaptation makes the network more susceptible to chaos, despite being a stabilizing negative feedback at the single-neuron level. The precise form of $g_c$ and the competition between static and oscillatory onset are the subject of Sec.~\ref{sec:spectral}.

\section{Spectral boundary and chaos onset}
\label{sec:spectral}
This section analyzes the first instability mechanism: loss of stability of the
coherent fixed point to the random bulk of the connectivity matrix. We derive a
frequency-dependent spectral boundary which sets the condition for the chaos threshold $g_c(\beta)$ and condition for a resonant chaos regime. 

The continuous set of eigenvalues generated by the random matrix $W$ can cross
the imaginary axis, while the discrete coherent eigenvalue generated by the
rank-one structure can cross independently. We treat the random eigenvalues in
this section and defer the coherent eigenvalue to Sec.~\ref{sec:coherent}.

Linearizing Eqs.~\eqref{eq:model_x}--\eqref{eq:model_a} around the fixed point, small perturbations $(\delta x_i, \delta a_i) \propto e^{st}$ satisfy
\begin{align}
(s\tau_m + 1)\,\delta x_i + \delta a_i &= \sum_j J_{ij}\,c_j\,\delta x_j, \label{eq:lin_x}\\
(s\tau_a + 1)\,\delta a_i &= \beta c_i\,\delta x_i, \label{eq:lin_a}
\end{align}
where $c_i = \sech^2(x_i^*)$ is the local gain~\eqref{eq:local_gain_def}. Substituting~\eqref{eq:lin_a} into~\eqref{eq:lin_x} to eliminate $\delta a_i$:
\begin{equation}
D_i(s)\,\delta x_i = \sum_j J_{ij}\,c_j\,\delta x_j,
\label{eq:reduced_eigen}
\end{equation}
where
\begin{equation}
D_i(s) \equiv s\tau_m + 1 + \frac{\beta c_i}{s\tau_a + 1}
\label{eq:Di}
\end{equation}
is the single-neuron characteristic polynomial (we also write $D(s;c)$ when treating the gain $c$ as a continuous variable rather than a neuron index). The $2N$-dimensional eigenvalue problem thereby reduces to the $N$-dimensional condition
\begin{equation}
\det\!\big[\diag(D_i(s)) - JC\big] = 0,
\label{eq:Ndim_chareq}
\end{equation}
where $C = \diag(c_j)$.

\begin{figure*}[t]
\centering
\includegraphics[width=\textwidth]{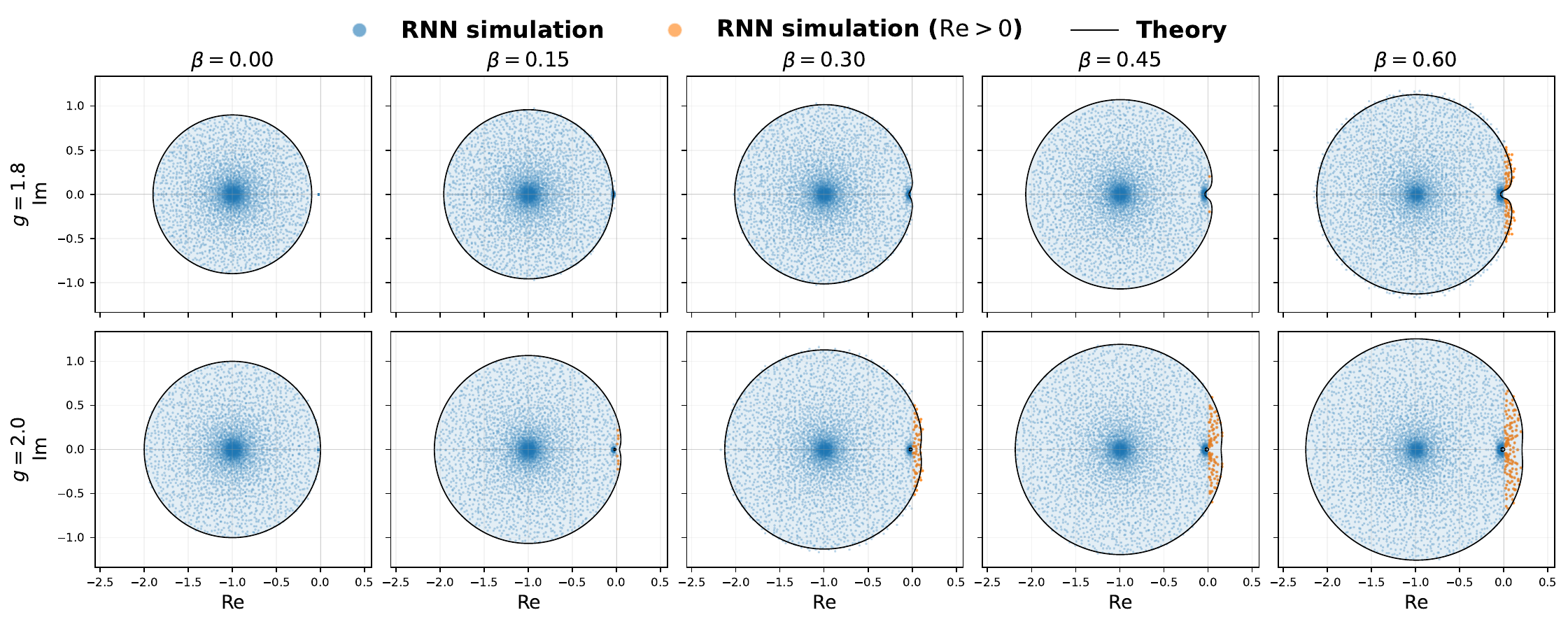}
\caption{\textbf{Adaptation deforms the random bulk toward a resonant
instability.}
The panels show eigenvalue spectra of the linearized dynamics in the complex
$s$-plane.
Top row: $g = 1.8$; bottom row: $g = 2.0$.
Columns correspond to increasing $\beta$.
Blue dots: numerically computed eigenvalues of the $2N$-dimensional Jacobian ($N = 4000$).
Orange dots: eigenvalues with $\re(s) > 0$ (unstable).
Black curves: theoretical spectral boundary from Eq.~\eqref{eq:parametric}.
At $\beta = 0$ the boundary is a circle centered at $s = -1$ with radius $g\sqrt{\chi_{2,x}}$.
As $\beta$ increases, the boundary develops a resonant bulge near $\im(s) \sim 1/\sqrt{\tau_a}$ and compresses at low frequencies, in quantitative agreement with the mean-field prediction.
At large $\beta$, eigenvalues cross the imaginary axis preferentially near the resonant frequency.
Parameters as in Fig.~\ref{fig:neuron_activity}.}
\label{fig:eig_spectra}
\end{figure*}

Dropping the rank-one term, the eigenvalue condition for the random part becomes $\det[\diag(D_i(s)) - gWC/\sqrt{N}] = 0$. The eigenvalues of $\diag(1/D_i)\,WC/\sqrt{N}$ coincide with those of $W\,\diag(c_j/D_j)/\sqrt{N}$ by the identity $\mathrm{spec}(AB)\setminus\{0\} = \mathrm{spec}(BA)\setminus\{0\}$; the latter is an i.i.d.\ random matrix with diagonal weights, whose squared spectral radius is $\E[|c/D|^2]$ by the circular law~\cite{RajanAbbott2006}. The spectral radius of the random eigenvalues is therefore
\begin{equation}
\rho^2(s) = g^2\,\E\!\left[\frac{c^2}{|D(s;c)|^2}\right].
\label{eq:rho2}
\end{equation}
The locus $\rho^2(s) = 1$ defines the spectral boundary in the complex $s$-plane (Fig.~\ref{fig:eig_spectra}).

To obtain a closed-form boundary, we replace $c$ inside $D$ by the population mean $\bar{\chi}_x$ while keeping $\chi_{2,x} = \E[c^2]$ in the numerator:
\begin{equation}
\rho^2(s) \approx \frac{g^2\chi_{2,x}}{|D_{\mathrm{mf}}(s)|^2}, \qquad D_{\mathrm{mf}}(s) = s + 1 + \frac{\beta\bar{\chi}_x}{s\tau_a + 1}.
\label{eq:rho2_mf}
\end{equation}
This approximation treats the gain as deterministic in the denominator but retains its second moment in the numerator; it is accurate when the variance of $c$ across the population is small compared to $\bar{\chi}_x^2$, which holds well away from the chaos threshold when most neurons are either deeply saturated or operating near the center of $\tanh$. Near onset, where the gain distribution is broadest, the full quenched expression~\eqref{eq:rho2} should be used.

The mean-field transfer function $\hat{G}(s) \equiv 1/D_{\mathrm{mf}}(s)$ is
\begin{equation}
\hat{G}(s) = \frac{s\tau_a + 1}{(s+1)(s\tau_a+1) + \beta\bar{\chi}_x}.
\label{eq:Ghat}
\end{equation}
We also write $\hat{G}_i(s) = 1/D_i(s)$ for the per-neuron transfer function, and $\hat{G}(s;c) = 1/D(s;c)$ when the gain is treated as a continuous parameter; the mean-field version is $\hat{G}(s) = \hat{G}(s;\bar{\chi}_x)$. On the imaginary axis,
\begin{equation}
|\hat{G}(i\omega)|^2 = \frac{1 + \omega^2\tau_a^2}{\omega^4\tau_a^2 + \omega^2(1 + \tau_a^2 - 2\beta\bar{\chi}_x\tau_a) + (1+\beta\bar{\chi}_x)^2}.
\label{eq:Ghat_sq}
\end{equation}
The eigenvalues of the single-neuron linearized dynamics are the roots of $D_{\mathrm{mf}}(s) = 0$. For $\beta\bar{\chi}_x > (\tau_a - 1)^2/(4\tau_a)$, these eigenvalues form a complex conjugate pair and the neuron exhibits damped oscillations at a frequency near $\omega_0 \sim 1/\sqrt{\tau_a}$.

Setting $\rho^2(s)=1$ in~\eqref{eq:rho2_mf} and parameterizing $\mu = g\sqrt{\chi_{2,x}}\,e^{i\theta}$ yields the quadratic
\begin{equation}
\tau_a\,s^2 + (\tau_a + 1 - \mu\tau_a)\,s + (1 + \beta\bar{\chi}_x - \mu) = 0.
\label{eq:parametric}
\end{equation}
As $\theta$ sweeps $[0,2\pi)$, the two roots trace a closed curve in the complex $s$-plane. Without adaptation this curve is the standard circle of radius $g\sqrt{\chi_{2,x}}$ centered at $s=-1$. With adaptation, the curve bulges outward near the resonant frequency $\omega_0$ and is compressed at low frequencies. The rightmost point determines the chaos threshold
\begin{equation}
g_c = \frac{1}{\sqrt{\chi_{2,x}\,\max_\omega |\hat{G}(i\omega)|^2}}.
\label{eq:gc}
\end{equation}

When $\beta = 0$, $|\hat{G}|^2 = 1/(1+\omega^2)$ peaks at $\omega = 0$ and gives $g_c^{(0)} = 1/\sqrt{\chi_{2,x}}$. For nonzero $\beta$, two competing behaviors emerge. At $\omega = 0$ the adaptation feedback suppresses the gain to $|\hat{G}(0)|^2 = 1/(1+\beta\bar{\chi}_x)^2$, which decreases with $\beta$. At intermediate frequencies the adaptation variable cannot fully follow the membrane potential, partially cancelling the low-pass filtering and producing a resonant peak at $\omega_* > 0$. The peak exists for essentially any nonzero $\beta$ when $\tau_a \gg 1$ (Appendix~\ref{app:spectral}). Writing $b \equiv \beta\bar{\chi}_x$, the two onset values are
\begin{align}
g_{\mathrm{stat}} &= \frac{1+b}{\sqrt{\chi_{2,x}}}, \label{eq:gstat}\\
g_{\mathrm{osc}} &= \frac{1}{\sqrt{\chi_{2,x}}}\sqrt{\frac{\tau_a + \Delta}{\tau_a}}\,, \label{eq:gosc}
\end{align}
where $\Delta(b) = 2[\sqrt{b(2+b)}-b]$.
Since $\Delta \leq 2$ for all $b$, the oscillatory threshold is
bounded above by $\sqrt{(\tau_a{+}2)/\tau_a}\,/\sqrt{\chi_{2,x}}$,
which for large $\tau_a$ lies only slightly above
$1/\sqrt{\chi_{2,x}}$.
The static threshold $(1+b)/\sqrt{\chi_{2,x}}$, by contrast,
grows without bound, so sufficiently strong adaptation always
makes the oscillatory onset lower. The corresponding deformation of the spectral boundary is visible
in Fig.~\ref{fig:eig_spectra}: at $\beta = 0$ the contour is a
circle, but increasing $\beta$ compresses it near the real axis
while producing an outward bulge near the resonant frequency.
The peak frequency scales as
$\omega_* \sim [b(2{+}b)]^{1/4}/\sqrt{\tau_a}$
(Appendix~\ref{app:spectral}).
This is the mechanism underlying the resonant chaos described by
Muscinelli et al.~\cite{MGS19}.

\section{Self-consistent fluctuation spectrum}
\label{sec:spectrum}

The previous section identifies where the random bulk first becomes unstable.
We now describe the stationary chaotic fluctuations above this boundary. The
goal is to compute the self-consistent firing-rate spectrum $S_\phi(f)$ around
the coherent fixed point, because this spectrum determines both the incoherent
single-neuron variability and the finite-size forcing that can drive the
coherent mode in Regime~II.

Above the chaos threshold $g > g_c$, each neuron fluctuates around its
fixed-point value $x_i^*$. We write
$\delta x_i(t) \equiv x_i(t)-x_i^*$ and compute the fluctuation spectrum using
a linear-response closure. In this closure, adaptation enters through
the local single-neuron transfer function, while the nonlinear firing-rate
covariance is closed using a Hermite expansion.

In the $N\to\infty$ DMFT~\eqref{eq:dmft_x}--\eqref{eq:dmft_a}, the overlap
$\kappa$ self-averages to $\kappa^*$ when the coherent fixed point is stable,
so the quenched input $h_i = m_i\kappa^* + g\eta_i^{(0)}$ is set by the
fixed-point value. The fluctuating part of the input to each neuron is
$g\,\delta\eta_i(t)$, where $\delta\eta_i(t) \equiv \eta_i(t) - \eta_i^{(0)}$
is the centered noise with power spectral density $S_\phi(f)$, the centered
firing-rate spectrum defined in Sec.~\ref{sec:dmft}. Each neuron receives an
independent realization of this process, but with the same spectral density
$S_\phi(f)$. (Coherent-mode fluctuations, which are $O(1/\sqrt{N})$, are
analyzed separately in Sec.~\ref{sec:coherent}.) Linearizing around the fixed
point and eliminating the adaptation perturbation $\delta a_i$
(Appendix~\ref{app:hermite}), each neuron acts as a linear filter of
$S_\phi(f)$. The resulting per-neuron membrane-potential fluctuation spectrum is 
\begin{equation}
S_{\delta x,i}(f) = g^2\,|\hat{G}_i(i\omega)|^2\,S_\phi(f), \qquad \omega = 2\pi f,
\label{eq:Sdx}
\end{equation}
where $\hat{G}_i(i\omega) = 1/D_i(i\omega)$ is the single-neuron transfer function~\eqref{eq:Di} evaluated at the local gain $c_i = \sech^2(x_i^*)$. The variance is $\sigma_{x,i}^2 = 2g^2\int_0^\infty |\hat{G}_i|^2 S_\phi\,\dd f$ and the normalized autocorrelation is $r_i(\tau) = C_{\delta x,i}(\tau)/\sigma_{x,i}^2$.

The DMFT self-consistency~\eqref{eq:selfcons} requires the population-averaged firing-rate autocorrelation $R_\phi(\tau)$ to match the single-neuron statistics. Since $\delta x_i$ is Gaussian, the nonlinear output $\phi(x_i^* + \delta x_i)$ can be expanded in the probabilist Hermite polynomials $\He_p$ ($\He_0 = 1$, $\He_1 = z$, $\He_2 = z^2 - 1$, \ldots), which form an orthogonal basis with respect to the standard Gaussian measure. The Mehler formula states that for a stationary unit-variance Gaussian process $z(t)$ with autocorrelation $r(\tau)$, the cross-moments satisfy $\E[\He_p(z(t))\He_q(z(t+\tau))] = \delta_{pq}\,p!\,r(\tau)^p$, so that different Hermite orders are uncorrelated at all lags. Applying this (Appendix~\ref{app:hermite}), the per-neuron firing-rate autocovariance takes the form
\begin{equation}
C_{\phi,i}(\tau) = \sum_{p=1}^P \frac{(b_p^{(i)})^2}{p!}\,r_i(\tau)^p,
\label{eq:Cphi_i}
\end{equation}
where $b_p^{(i)} = \E_z[\tanh(x_i^* + \sigma_{x,i}z)\,\He_p(z)]$ are the Hermite coefficients and $P$ is the truncation order. The $p = 1$ term is the linearized contribution (proportional to $r$ itself). The higher-order terms generate spectral harmonics and broadening.

\begin{figure*}[t]
\centering
\includegraphics[width=\textwidth]{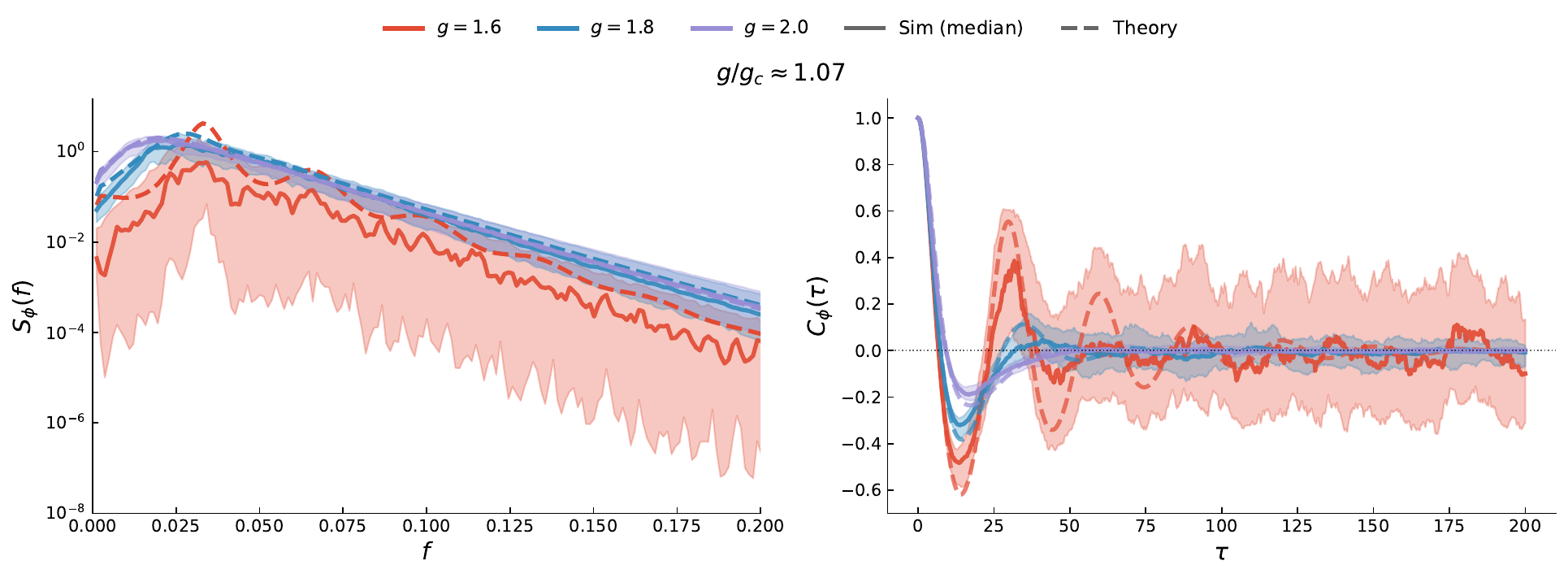}
\caption{\textbf{DMFT captures the adaptation-induced resonant
peak.}
The left panel shows the firing-rate power spectral density $S_\phi(f)$ just
above the chaos threshold, at approximately $g/g_c \approx 1.07$. The right
panel shows the corresponding autocovariance function $C_\phi(\tau)$ for the
same parameter values. Colors indicate $g = 1.6$, $1.8$, and $2.0$. Solid
curves show simulation medians, and shaded bands show interquartile ranges.
Dashed curves show the self-consistent spectral DMFT prediction from
Sec.~\ref{sec:spectrum}. The spectral peak near $f\approx 0.02$--$0.03$
reflects the resonant frequency introduced by adaptation. Parameters are as in
Fig.~\ref{fig:neuron_activity}.}
\label{fig:spectra_ggc}
\end{figure*}

\begin{figure}[!htbp]
\centering
\includegraphics[width=\columnwidth]{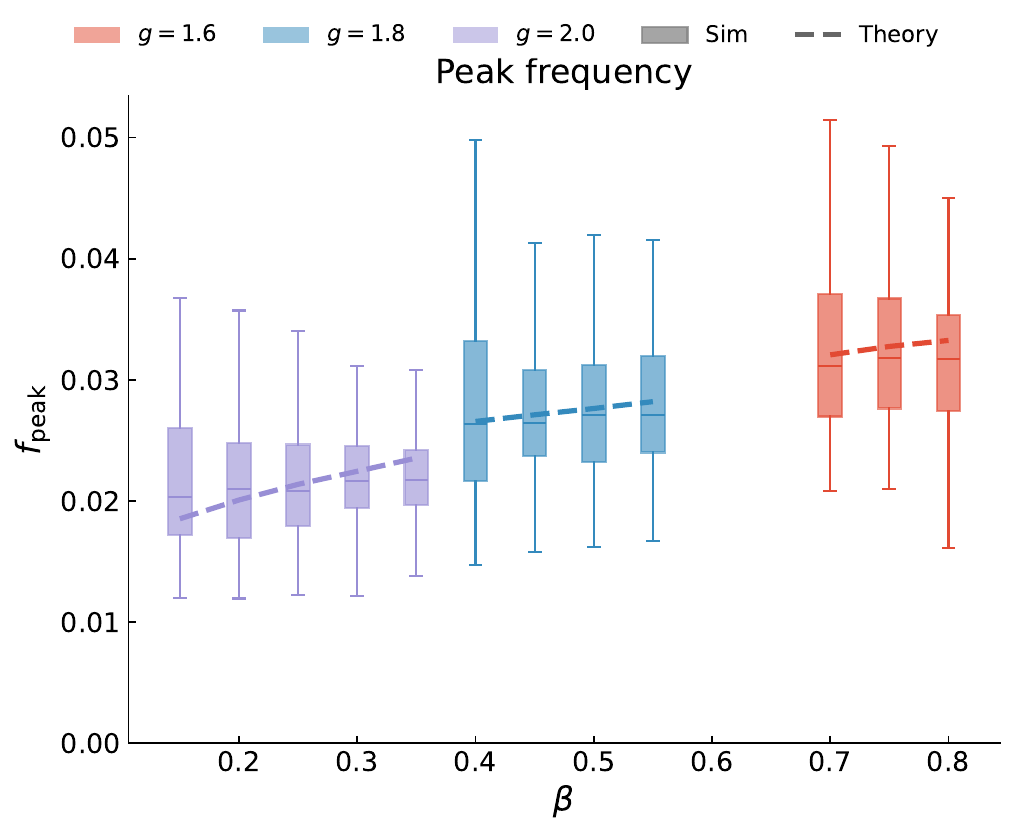}

\caption{\textbf{The resonant peak shifts upward with adaptation strength.}
Peak frequency $f_{\mathrm{peak}}$ of the fluctuation spectrum versus $\beta$ for $g = 1.6$ (red), $1.8$ (blue), $2.0$ (purple).
Box plots: simulation (median, interquartile range, and extremes across neurons).
Dashed lines: DMFT prediction.
The peak frequency increases with $\beta$ as the adaptation resonance shifts to higher frequencies.
Parameters as in Fig.~\ref{fig:neuron_activity}.}
\label{fig:fpeak}
\end{figure}

\begin{figure}[!htbp]
\centering
\includegraphics[width=\columnwidth]{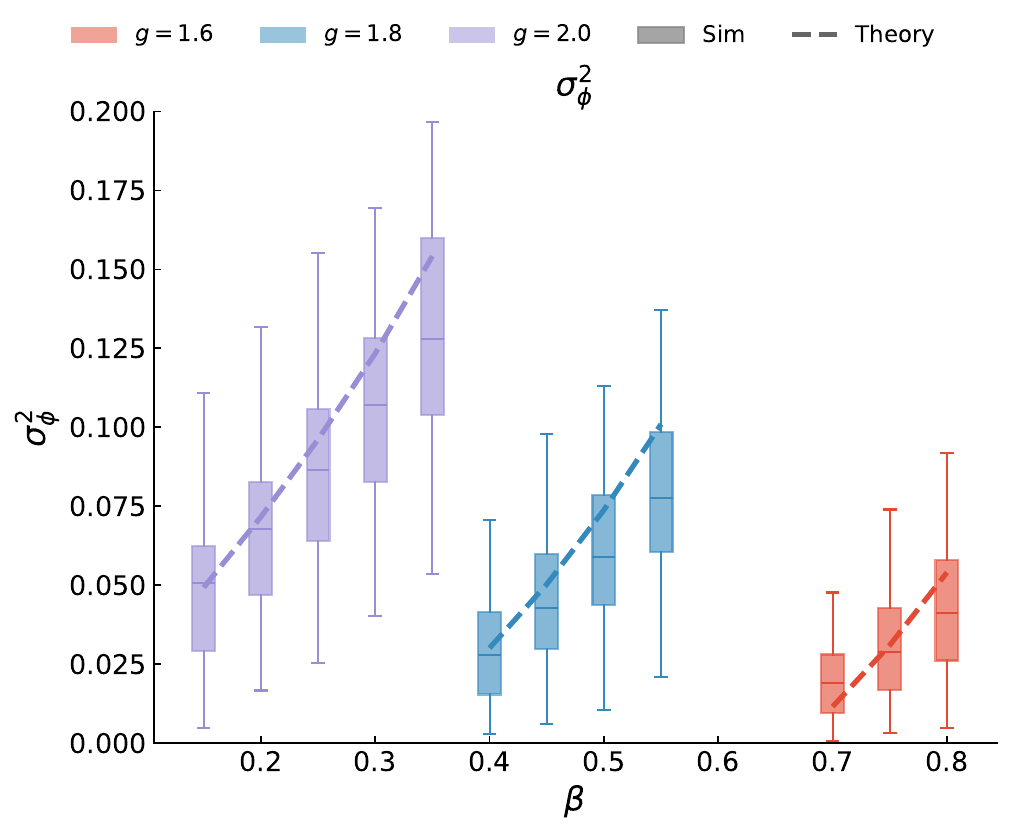}
\caption{\textbf{Adaptation increases incoherent firing-rate variance above
chaos.}
Incoherent variance $\sigma_\phi^2$ of firing-rate fluctuations versus $\beta$.
Box plots: simulation; dashed lines: DMFT theory.
Colors as in Fig.~\ref{fig:fpeak}.
The variance increases with $\beta$ within each $g$ series, reflecting stronger chaotic fluctuations as adaptation modifies the gain distribution.
Parameters as in Fig.~\ref{fig:neuron_activity}.}
\label{fig:variance}
\end{figure}

All neuron-specific quantities are deterministic functions of the quenched input $h_i$. The fixed point $x^* = F^{-1}(h)$ determines the local gain $c = \sech^2(x^*)$, which sets the transfer function $|\hat{G}(i\omega;c)|^2$ and hence the fluctuation spectrum $S_{\delta x}$, variance $\sigma_x$, normalized autocorrelation $r$, and Hermite coefficients $b_p$. Since $h \sim \mathcal{N}(0,\Sigma_h^2)$, the population-averaged autocovariance $C_\phi(\tau) = \E_h[C_{\phi,i}(\tau)]$ reduces to a one-dimensional Gaussian integral over $h$. Its one-sided cosine transform $S_\phi'(f) = 2\int_0^\infty C_\phi(\tau)\cos(2\pi f\tau)\,d\tau$ yields a new spectrum, defining the spectral map $\mathcal{T}: S_\phi \mapsto S_\phi'$. The self-consistent spectrum $S_\phi^* = \mathcal{T}[S_\phi^*]$ is found by damped iteration. The incoherent variance $\sigma_\phi^2 \equiv C_\phi(0) = 2\int_0^\infty S_\phi^*(f)\,df$ measures the total power in the firing-rate fluctuations. Figure~\ref{fig:spectra_ggc} compares the theoretical spectrum against simulations, and Figs.~\ref{fig:fpeak} and~\ref{fig:variance} show the peak frequency and $\sigma_\phi^2$ versus $\beta$.

Below chaos, $\mathcal{T}[0] = 0$ is the unique fixed point. Linearizing the map around $S_\phi = 0$, only the $p=1$ Hermite coefficient survives and the linearized map acts as frequency-by-frequency multiplication by $\rho^2(i\omega)$ (Appendix~\ref{app:hermite}). A nontrivial spectral fixed point therefore bifurcates from zero precisely when $\max_\omega \rho^2(i\omega) = 1$, confirming that the eigenvalue analysis of Sec.~\ref{sec:spectral} and the spectral self-consistency give the same chaos threshold.

Above chaos, chaotic fluctuations modify the effective gains. The fluctuation-averaged gain is
\begin{equation}
\tilde{c}(h) = \E_\zeta\!\left[\sech^2\!\big(F^{-1}(h) + \sigma_x(h)\zeta\big)\right], \qquad \zeta\sim\mathcal{N}(0,1),
\label{eq:eff_gain}
\end{equation}
with $\sigma_x(h)$ the local fluctuation amplitude, determined self-consistently from $\sigma_x^2(h) = 2g^2\int_0^\infty |\hat{G}(i\omega;c(h))|^2\,S_\phi^*\,\dd f$. The self-consistency~\eqref{eq:selfcons_fp} generalizes to (Appendix~\ref{app:fluct_corrections})
\begin{equation}
\frac{1}{\lambda_{\mathrm{eff}}} \approx \E_h\!\left[\frac{\tilde{c}(h)}{1+\beta\,c(h)}\right],
\label{eq:eff_selfcons}
\end{equation}
to leading order, where $c(h) = \sech^2(F^{-1}(h))$ is the unperturbed local gain~\eqref{eq:local_gain_def}. Below chaos, $\tilde{c}(h) = c(h)$ and this recovers~\eqref{eq:selfcons_fp}.

\section{Coherent mode stability and phase diagram}
\label{sec:coherent}

Sections~\ref{sec:spectral} and~\ref{sec:spectrum} analyzed the bulk
eigenvalues from the random connectivity. We now turn to the discrete
eigenvalue associated with the rank-one structure.
Restoring the rank-one component to~\eqref{eq:Ndim_chareq} gives a
rank-one outlier condition for coherent perturbations. A perturbation of the
coherent overlap changes the input to neuron $i$ by
\begin{equation}
\delta \hat I_i(s) = m_i\,\delta\hat\kappa(s).
\end{equation}
For a neuron with local linear gain $c$, the linearized adaptation dynamics
give
\begin{align}
\delta\hat\phi_i(s)
&=
c\,\hat{G}(s;c)\,\delta\hat I_i(s),
\label{eq:single_neuron_response}\\
\hat{G}(s;c)
&=
\frac{s\tau_a+1}
     {(s+1)(s\tau_a+1)+\beta c}.
\label{eq:single_neuron_transfer}
\end{align}
Equivalently,
$\hat{G}(s;c)=[s+1+\beta c/(s\tau_a+1)]^{-1}$ is the
membrane-potential response after eliminating the adaptation variable.

The gain entering this linear response is the fixed-point gain below the
chaos threshold and an effective fluctuation-averaged gain above it. We write
\begin{equation}
c_{\mathrm{lin}}(h) =
\begin{cases}
c(h), & \text{below the chaos threshold},\\[1mm]
\tilde{c}(h), & \text{above the chaos threshold},
\end{cases}
\label{eq:clin_def}
\end{equation}
where $c(h)=\sech^2(F^{-1}(h))$ and $\tilde{c}(h)$ is defined in
Eq.~\eqref{eq:eff_gain}. Above chaos, this replacement is an effective
linear-response approximation in which temporal fluctuations are represented
by a fluctuation-averaged local gain.

Projecting the firing-rate perturbation back onto the output loading gives
\begin{equation}
\delta\hat\kappa(s)=L_N(s)\,\delta\hat\kappa(s),
\label{eq:finiteN_feedback}
\end{equation}
where the finite-$N$ diagonal loop gain is
\begin{equation}
\begin{split}
L_N(s)
&=
\frac{1}{N}\sum_i
n_i m_i\,c_{\mathrm{lin}}(h_i) \\
&\quad\times
\hat{G}\!\big(s;c_{\mathrm{lin}}(h_i)\big).
\end{split}
\label{eq:LN_def}
\end{equation}
Thus a nonzero coherent perturbation satisfies
\begin{equation}
L_N(s)=1.
\label{eq:coherent_finiteN}
\end{equation}

For an outlier eigenvalue outside the random bulk, $\rho^2(s)<1$, the
off-diagonal terms of the resolvent are $O(N^{-1/2})$, so the diagonal loop
gain has a deterministic large-$N$ limit (Appendix~\ref{app:coherent}):
\begin{equation}
\begin{split}
L_N(s)
&\xrightarrow{N\to\infty}
\E\!\left[
nm\,c_{\mathrm{lin}}(h)\,
\hat{G}\!\big(s;c_{\mathrm{lin}}(h)\big)
\right].
\end{split}
\label{eq:coherent_limit}
\end{equation}
The expectation is over the joint Gaussian distribution of $(n,m,h)$.
Conditioning on the quenched input $h$ gives
\begin{equation}
\E[nm\mid h]
=
\lambda_{\mathrm{eff}}\big((1-\alpha)+\alpha w^2\big),
\qquad
w=\frac{h}{\Sigma_h},
\end{equation}
where
\begin{equation}
\alpha =
\frac{\sigma_m^2\kappa^2}{\Sigma_h^2},
\qquad
\Sigma_h^2 = \sigma_m^2\kappa^2 + g^2 Q.
\end{equation}
Here $\alpha$ is the fraction of the total input variance contributed by the
rank-one structured component. Applying the law of iterated expectations to
Eq.~\eqref{eq:coherent_limit} gives the coherent loop gain
\begin{equation}
\begin{split}
L(s)
&=
\lambda_{\mathrm{eff}}\,
\E\!\left[
\big((1-\alpha)+\alpha w^2\big)\,
c_{\mathrm{lin}}(h) \right. \\
&\qquad\left.
\times\hat{G}\!\big(s;c_{\mathrm{lin}}(h)\big)
\right].
\end{split}
\label{eq:loop_tf}
\end{equation}
The rank-one outlier eigenvalues are the solutions of
\begin{equation}
L(s)=1.
\label{eq:coherent_eig_condition}
\end{equation}

When $\beta=0$, the adaptation variable decouples from the membrane dynamics
and $\hat{G}(s;c)=1/(s+1)$ is independent of $c$. The outlier condition then
gives a purely real eigenvalue,
\begin{equation}
s =
-1
+
\lambda_{\mathrm{eff}}\,
\E\!\left[
\big((1-\alpha)+\alpha w^2\big)c_{\mathrm{lin}}(h)
\right].
\end{equation}
Thus a real rank-one structured mode can change the stability of coherent
fixed points, but it does not by itself provide the phase lag needed for an
oscillatory coherent instability. This is consistent with the standard
rank-one result that real rank-one structure produces fixed-point dynamics,
whereas sustained oscillations require additional structured dimensions with
appropriate geometry~\cite{MO18}. For $\beta>0$, adaptation introduces the
frequency-dependent factor $\hat{G}(i\omega;c)$, which can give the coherent
feedback loop a nonzero phase lag and allows the rank-one outlier to cross the
imaginary axis through a Hopf bifurcation.

\subsection{Hopf bifurcation and phase diagram}
We next determine when the coherent fixed point loses stability and combine this boundary with the random-bulk chaos boundary.

At zero frequency, Eq.~\eqref{eq:single_neuron_transfer} gives
$\hat{G}(0;c)=1/(1+\beta c)$. The static coherent loop gain is therefore
\begin{equation}
L(0) =
\lambda_{\mathrm{eff}}\,
\E\!\left[
  \frac{(1-\alpha+\alpha w^2)c_{\mathrm{lin}}(h)}
       {1+\beta c_{\mathrm{lin}}(h)}
\right].
\label{eq:L0}
\end{equation}
Below the chaos threshold, $c_{\mathrm{lin}}(h)=c(h)$. Comparing
Eq.~\eqref{eq:L0} with the fixed-point condition~\eqref{eq:selfcons_fp} gives
\begin{equation}
L(0)-1 =
\lambda_{\mathrm{eff}}\alpha\,
\E\!\left[
  (w^2-1)\frac{c(h)}{1+\beta c(h)}
\right].
\label{eq:L0_minus_one}
\end{equation}
The projection factor $(1-\alpha+\alpha w^2)$ appears because a coherent
perturbation is injected through $m_i$ and read out through $n_i$. Although
this factor has unit mean, it is generally correlated with the static gain
$c(h)/(1+\beta c(h))$. Thus Eq.~\eqref{eq:selfcons_fp} determines the
nonlinear coherent fixed point, whereas Eq.~\eqref{eq:coherent_eig_condition}
determines the linear eigenvalues around that fixed point. The condition
$L(0)=1$ corresponds to a static coherent instability.

The oscillatory coherent instability occurs when a complex-conjugate pair of
rank-one outlier eigenvalues crosses the imaginary axis. The Hopf boundary is
therefore determined by a frequency $\omega^*>0$ satisfying
\begin{equation}
\im\,L(i\omega^*) = 0, \qquad \re\,L(i\omega^*) = 1.
\label{eq:hopf}
\end{equation}
The first condition pins the frequency; the second determines the critical
adaptation strength $\beta_H$. Physically, the adaptation adds positive phase to $\hat{G}(i\omega;c)$
through the numerator factor $(i\omega\tau_a + 1)$. Without adaptation,
$\im\,L(i\omega) < 0$ for all $\omega > 0$ and no Hopf bifurcation is
possible. With adaptation, the positive phase overcomes the negative
phase from the membrane dynamics at a frequency
$\omega^* \sim 1/\sqrt{\tau_a}$, producing a zero crossing of
$\im\,L$. As $\beta$ increases, $\re\,L(i\omega^*)$ grows until it
exceeds unity and the coherent mode goes unstable. Below chaos, $\Sigma_h^2$ and the gains do not depend on $g$, so
$\beta_H$ is approximately independent of $g$ and the Hopf boundary is nearly
horizontal in the $(\beta,g)$ plane. Above chaos, $\tilde{c}$ and $\Sigma_h^2$
depend on $g$ through the spectral DMFT, and the Hopf boundary curves.

\begin{figure*}[t]
\centering
\includegraphics[width=\textwidth]{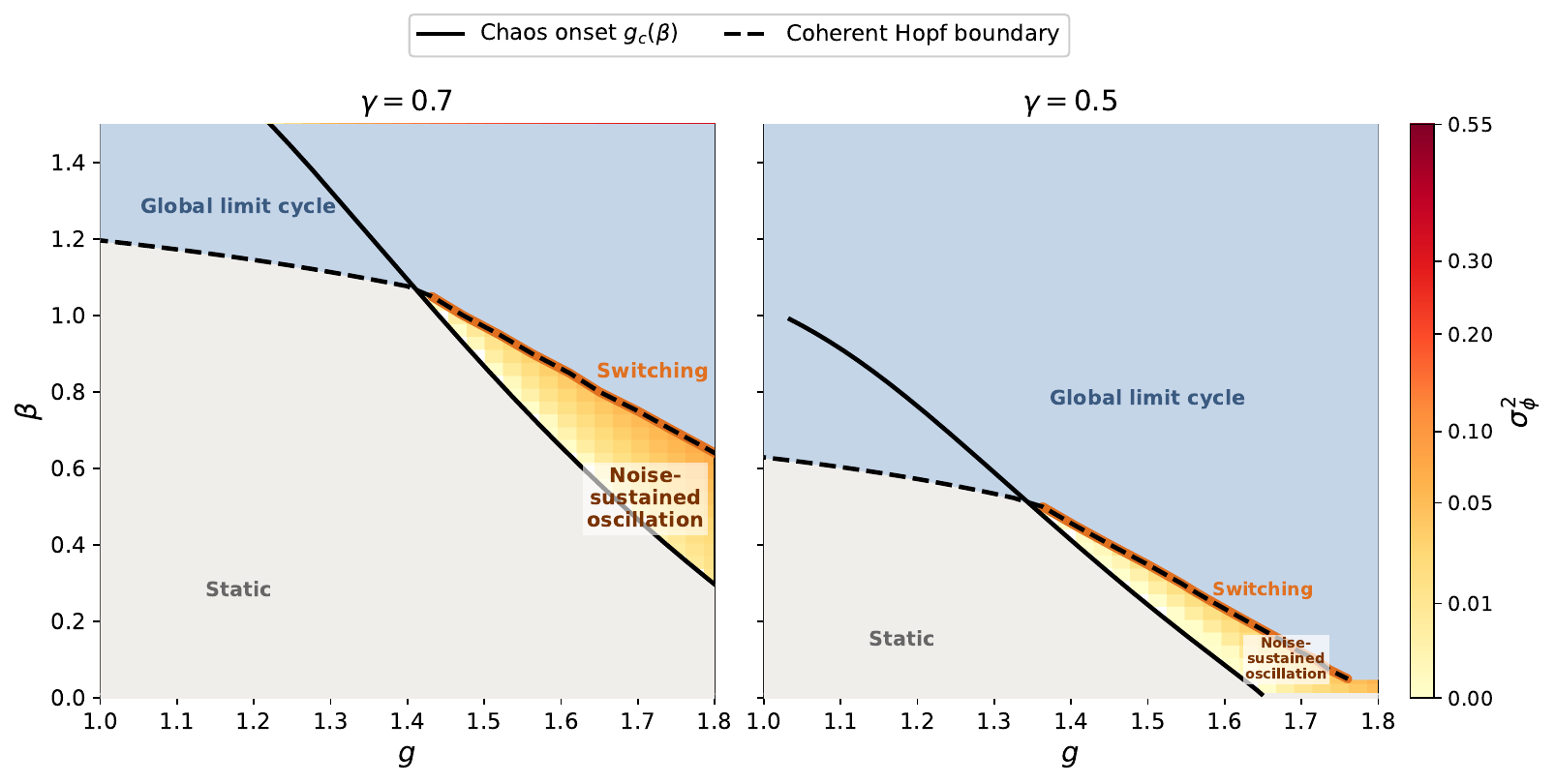}
\caption{\textbf{Chaos and coherent Hopf boundaries partition the phase
diagram.} Phase diagram in the $(g, \beta)$ plane for $\gamma = 0.7$ (left) and
$\gamma = 0.5$ (right).
Solid line: chaos onset $g_c(\beta)$ from Eq.~\eqref{eq:gc}.
Dashed line: coherent Hopf boundary from Eq.~\eqref{eq:hopf}.
Grey region: static coherent state.
Between the two boundaries, the color map shows the incoherent variance
$\sigma_\phi^2$ from the spectral DMFT. This is the noise-sustained oscillation
regime, where the coherent mode is a stable focus driven by chaotic
fluctuations. The orange segment of the Hopf boundary marks the switching
region, where the two boundaries are close and chaotic fluctuations
can drive transitions between basins. Blue region: global limit cycle.}
\label{fig:phase_diagram}
\end{figure*}

Together, the chaos boundary $g_c(\beta)$ and the Hopf boundary $\beta_H(g)$,
along with the existence condition $\beta < \lambda_{\mathrm{eff}} - 1$,
partition the $(\beta,g)$ plane into four regions
(Fig.~\ref{fig:phase_diagram}). Regime~I (static) lies below both boundaries.
Regime~II (noise-sustained oscillation) lies above the chaos boundary but below
the Hopf boundary, where the coherent mode is a stable focus. In this region,
the chaotic fluctuations $S_\phi^*$ from Sec.~\ref{sec:spectrum} project onto
the coherent mode through the random connectivity.  The coherent mode's transfer
function $\hat{G}$ selectively amplifies the spectral peak near the adaptation
resonance, producing sustained overlap oscillations even though the mode is
linearly stable. The progression from regular to irregular oscillations
within Regime~II reflects the approach to $\beta_H$. As the stability margin
shrinks, the noise spectrum becomes broader, and the
oscillation becomes less coherent. Regime~III (switching) occurs near the Hopf
boundary, where fluctuations can drive the trajectory across the separatrix
between the two symmetric basins. Regime~IV (global oscillation) lies above the
Hopf boundary. By the $\mathbb{Z}_2$ symmetry of the model, both fixed points
$\pm\kappa^*$ lose stability simultaneously, and the resulting limit cycle
connects the two basins.

The order in which the two boundaries are crossed determines the route through
the phase diagram. For $g > g_c$ at small $\beta$, chaos sets in first and the
network passes through Regimes~II and~III before reaching the Hopf boundary
and entering Regime~IV. For $g < g_c$, the coherent mode loses stability
before chaos onset and the network jumps directly from the static state
(Regime~I) to the global limit cycle (Regime~IV), with no intermediate
noise-sustained regimes.

\section{Low-dimensional dynamics}
\label{sec:reduced}

\begin{figure*}[t]
\centering
\includegraphics[width=\textwidth]{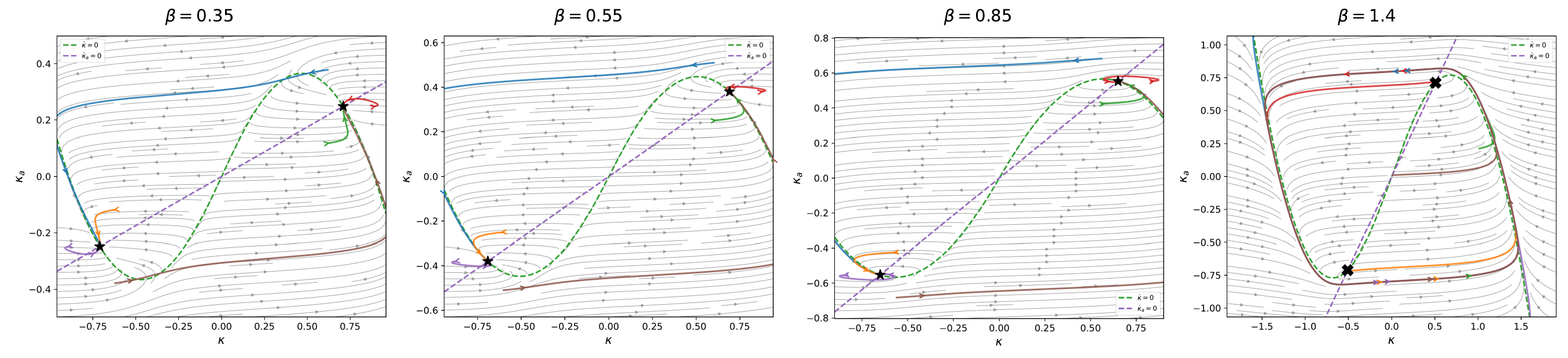}
\caption{\textbf{The reduced deterministic model captures the local
bifurcation structure.} Phase portraits of the reduced deterministic model~\eqref{eq:rd_3d} in the $(\kappa, \kappa_a)$ plane at $g = 1.8$ for increasing adaptation strength.
Green dashed curves: $\dot{\kappa} = 0$ nullcline.
Purple dashed lines: $\dot{\kappa}_a = 0$ nullcline ($\kappa_a = \beta\kappa$).
Stars: stable fixed points; crosses: unstable fixed points (where applicable).
Gray streamlines show the vector field; colored trajectories illustrate representative solutions from different initial conditions.
(a)~$\beta = 0.35$: the fixed points are stable nodes; trajectories converge monotonically.
(b)~$\beta = 0.55$: the fixed points have become stable foci; the spiraling return flow underlies the noise-sustained oscillation of Regime~II.
(c)~$\beta = 0.85$: the foci approach the Hopf boundary; the narrowed basins of attraction permit noise-driven switching between wells (Regime~III).
(d)~$\beta = 1.40$: the fixed points have lost stability and a global limit cycle encircles both wells (Regime~IV).
Parameters as in Fig.~\ref{fig:neuron_activity}.}
\label{fig:ode_phase_panel}
\end{figure*}

\begin{figure*}[t]
\centering
\includegraphics[width=0.32\textwidth]{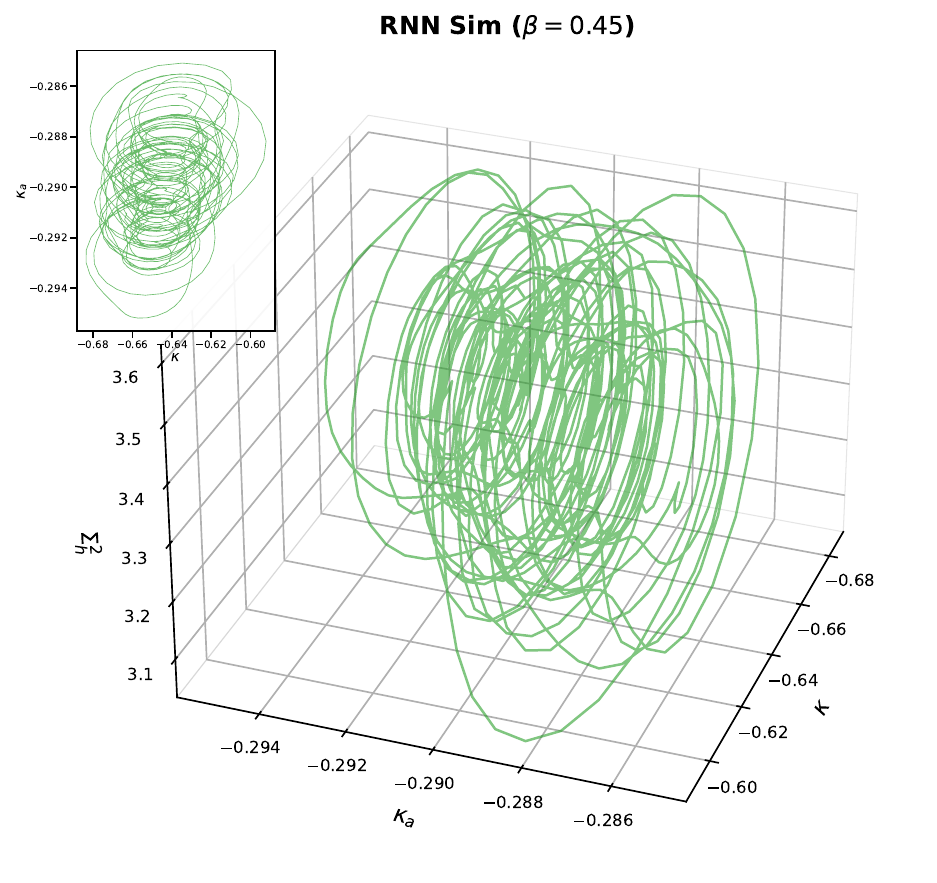}\hfill
\includegraphics[width=0.32\textwidth]{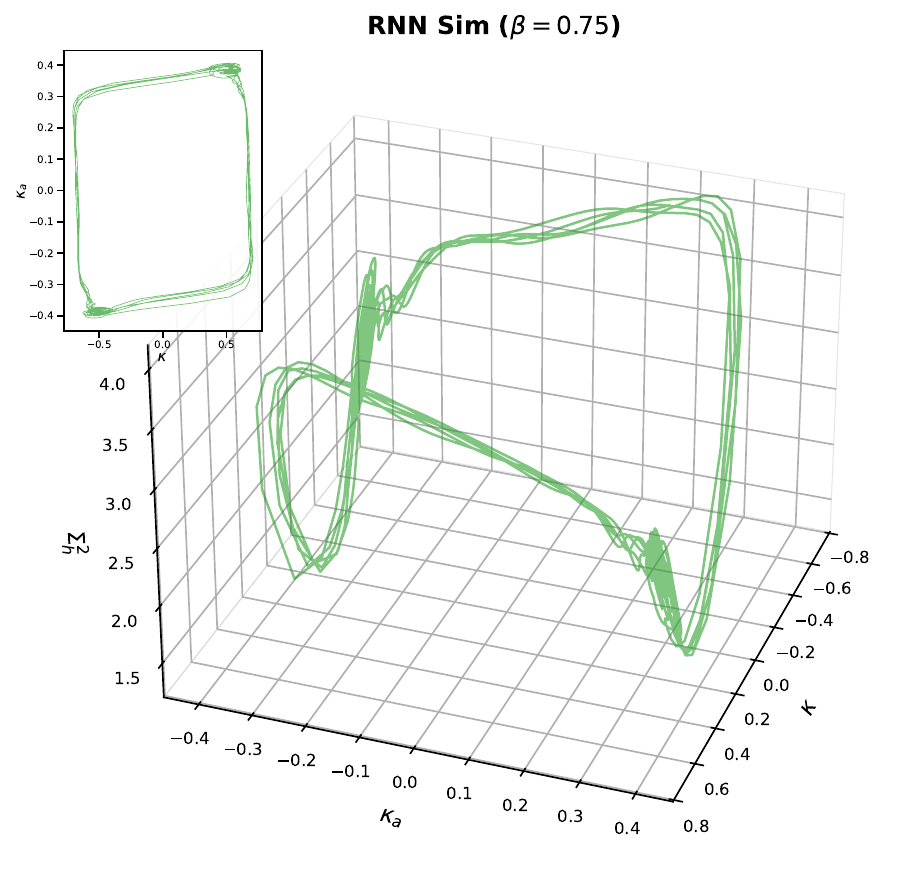}\hfill
\includegraphics[width=0.32\textwidth]{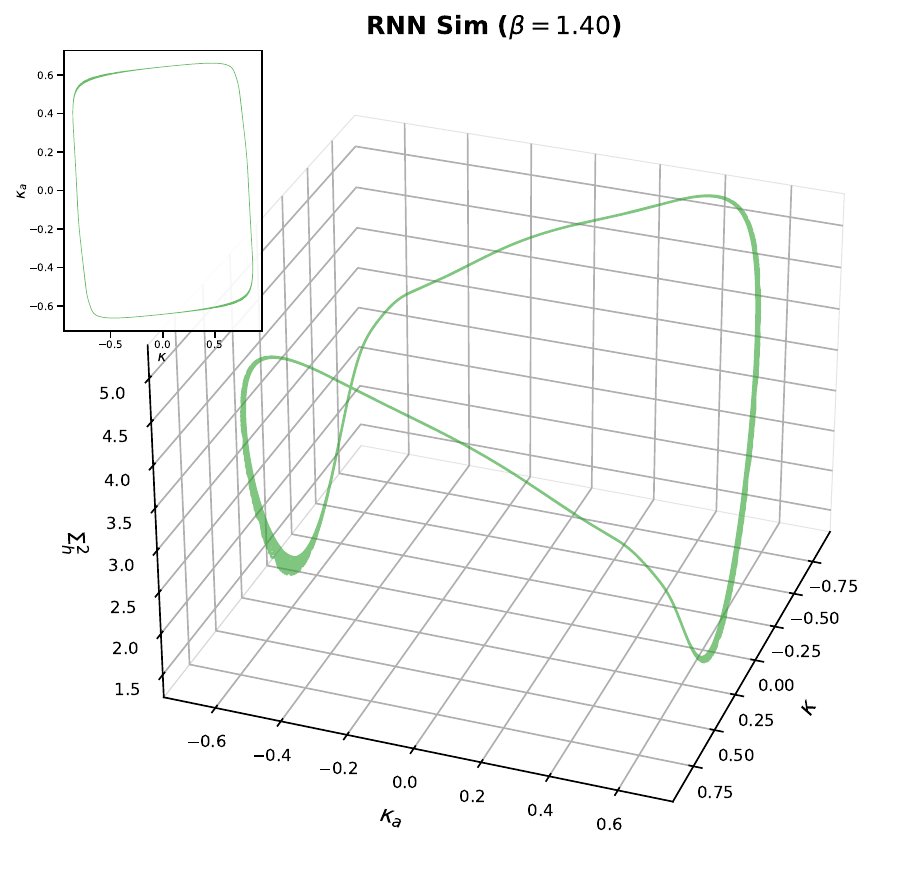}\\[3pt]
\includegraphics[width=0.32\textwidth]{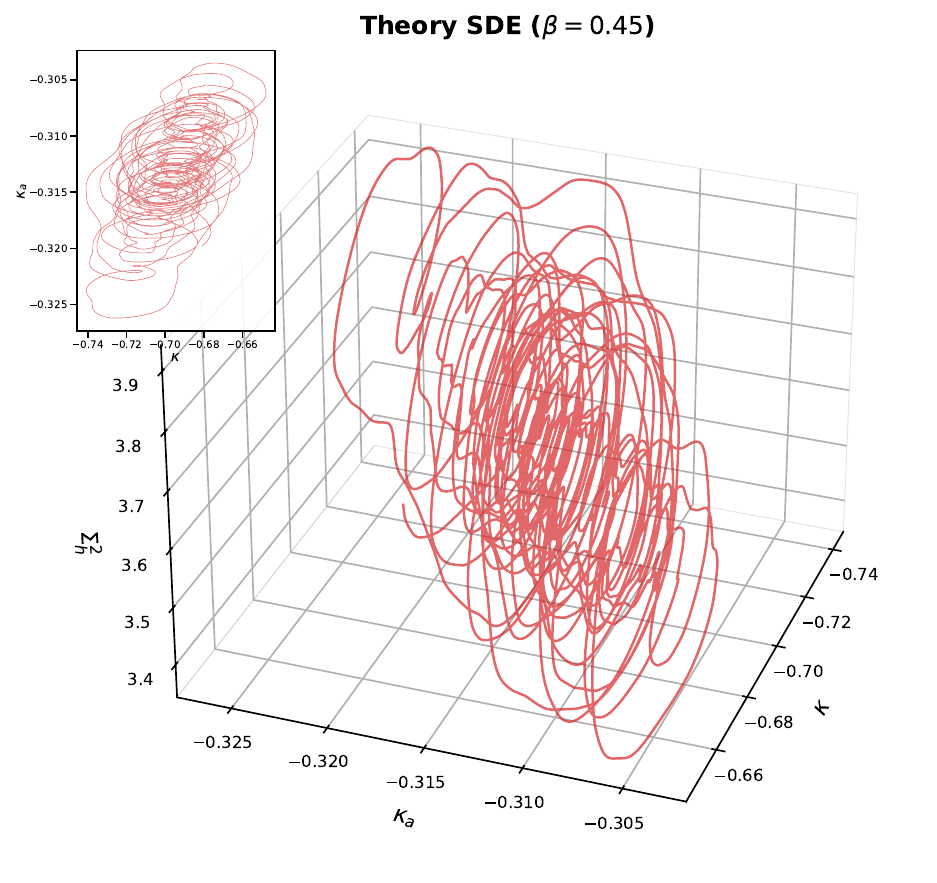}\hfill
\includegraphics[width=0.32\textwidth]{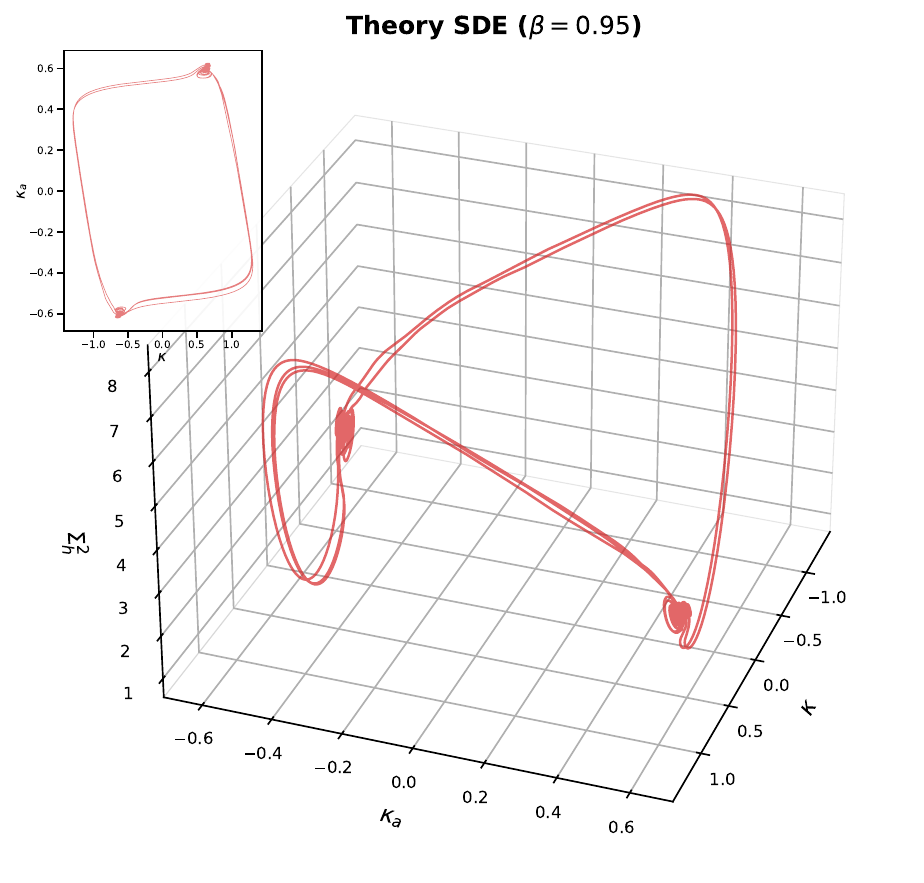}\hfill
\includegraphics[width=0.32\textwidth]{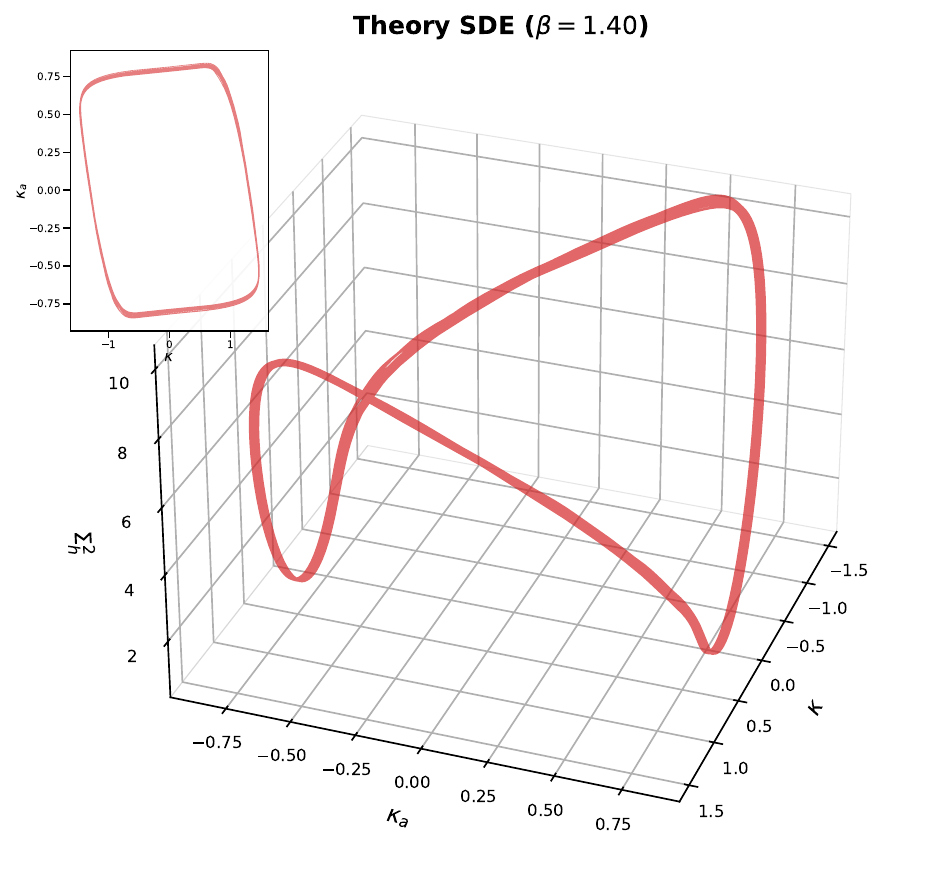}

\caption{\textbf{A stochastic reduced model reproduces the regime transitions and the global shape of the limit cycle.} Three-dimensional trajectories in $(\kappa, \kappa_a, \Sigma_h^2)$ space showing the progression of dynamical regimes.
Top row: full RNN simulations ($N = 4000$, $g = 1.8$).
Bottom row: low-dimensional SDE model~\eqref{eq:rd_SDE} with DMFT-calibrated parameters.
Left column ($\beta = 0.45$ for both): noise-sustained oscillation confined to a single well; the trajectory spirals around a stable focus, with the chaotic bulk providing the energy that maintains the oscillation.
Middle column ($\beta = 0.75$ for the RNN, $\beta = 0.95$ for the SDE): irregular switching between the two wells; chaotic fluctuations occasionally kick the trajectory across the separatrix.
Right column ($\beta = 1.40$ for both): global limit cycle carrying $\kappa$ between both wells. The reduced SDE model qualitatively reproduces the full sequence of regimes observed in the network simulations.
Parameters as in Fig.~\ref{fig:neuron_activity}.}
\label{fig:phase_3d_panel}
\end{figure*}

The DMFT of the preceding sections characterizes the fixed points and their
stability boundaries. To visualize the bifurcation sequence in the phase space
of the order parameters, we derive a reduced system for $\kappa(t)$,
$\kappa_a(t)$, and $Q(t)$ using the Gaussian closure approach of the low-rank
RNN literature~\cite{MO18,Schuessler2020}.

From Sec.~\ref{sec:model}, $\kappa_a$ satisfies the exact equation
$\tau_a\,\dot{\kappa}_a = -\kappa_a + \beta\,\kappa$. Closing the dynamics
for $\kappa$ requires approximation. Projecting the membrane-potential equation
onto the loading vectors $\bm{n}$ and applying an instantaneous Gaussian
closure (Appendix~\ref{app:reduced}) gives a relaxation equation for $\kappa$
with an effective gain
\begin{equation}
G_{\mathrm{eff}}(\Sigma_h^2, \beta) = \frac{\bar{\chi}_{\mathrm{eff}}}{1 - \beta\,\bar{\chi}_{\mathrm{eff}}},
\label{eq:Geff}
\end{equation}
which is constructed so that the reduced system reproduces the DMFT fixed-point
condition $\lambda_{\mathrm{eff}}\bar{\chi}_{\mathrm{eff}} = 1$, and reduces
to $\bar{\chi}_{\mathrm{eff}}$ at $\beta = 0$. The three-dimensional reduced
ODE is
\begin{equation}
\begin{aligned}
\tau_m\,\dot{\kappa} &= -\kappa + G_{\mathrm{eff}}(\Sigma_h^2,\beta)\,(\lambda_{\mathrm{eff}}\kappa - \kappa_a),\\
\tau_a\,\dot{\kappa}_a &= -\kappa_a + \beta\,\kappa,\\
\tau_c\,\dot{Q} &= -Q + Q_{\mathrm{fp}}(\Sigma_h^2,\beta),
\end{aligned}
\label{eq:rd_3d}
\end{equation}
with $\Sigma_h^2 = \sigma_m^2\kappa^2 + g^2Q$ and
$Q_{\mathrm{fp}}(\Sigma_h^2,\beta) = \E_Z[\tanh^2(F^{-1}(\Sigma_h Z))]$.
An estimate of the incoherent variance relaxation time is
\[
\tau_c \approx
\frac{\tau_{\mathrm{auto}}}
{2\left(1 - g^2 Q'_{\mathrm{fp}}(\Sigma_h^{2*},\beta)\right)},
\]
where $\tau_{\mathrm{auto}} = S_\phi^*(0)/(2C_\phi(0))$ is the DMFT autocorrelation time (Appendix~\ref{app:reduced}). 
When $\tau_c \ll \tau_m$, $Q$ relaxes rapidly to its instantaneous equilibrium
and the system reduces to a planar model for $(\kappa, \kappa_a)$ with
characteristic equation
\begin{equation}
(s\tau_m + 1 - \lambda_{\mathrm{eff}}G_{\mathrm{eff}})(s\tau_a + 1) + \beta G_{\mathrm{eff}} = 0,
\label{eq:rd_chareq}
\end{equation}
a mean-gain approximation to~\eqref{eq:loop_tf} in which the quenched
population average is replaced by evaluation at $G_{\mathrm{eff}}$.

\subsection{Stability progression with increasing adaptation}
We use the reduced deterministic model to visualize how the coherent fixed points evolve from stable nodes to stable foci and then to an unstable pair
surrounded by a limit cycle. 
Figure~\ref{fig:ode_phase_panel} shows the $(\kappa,\kappa_a)$ phase portraits for four values of $\beta$ at $g = 1.8$.

At $\beta = 0.35$ the $\dot{\kappa} = 0$ nullcline (green) intersects the
$\dot{\kappa}_a = \beta\kappa$ nullcline (purple) at two symmetric fixed
points $(\pm\kappa^*, \pm\kappa_a^*)$. The eigenvalues of the Jacobian at
these fixed points are real and negative. The fixed points are stable
nodes, and nearby trajectories converge monotonically. As $\beta$
increases to $0.55$, the eigenvalues acquire nonzero imaginary parts while
retaining negative real parts: the fixed points become stable foci. In
the presence of chaotic fluctuations (when $g > g_c$), this spiraling return
flow produces the noise-sustained oscillation. The vector field
shows that trajectories approaching the fixed point execute damped oscillations
whose amplitude and frequency grow with~$\beta$.

At $\beta = 0.85$ the foci are close to the Hopf boundary and the stability
margin is small. The phase portrait reveals that the basins of attraction of
the two wells have narrowed. In the full network, chaotic fluctuations can now kick the
trajectory across the separatrix, causing the system to linger and oscillate
around one fixed point before a sufficiently large noise excursion drives it to
the other basin, producing the irregular switching between two basins.

At $\beta = 1.40$ the fixed points have lost stability through a Hopf
bifurcation (marked by crosses in Fig.~\ref{fig:ode_phase_panel}d). A stable
limit cycle encircling both wells has appeared, carrying $\kappa$ through
large-amplitude oscillations between $+\kappa^*$ and $-\kappa^*$. The limit cycle persists even in
the absence of noise ($g < g_c$), confirming that it arises from a genuine
bifurcation rather than from stochastic forcing.

\subsection{Stochastic forcing from random fluctuations}
We now estimate how chaotic fluctuations project onto the coherent mode and sustain oscillations below the Hopf boundary.
Above chaos, random fluctuations inject finite-size forcing into the coherent
overlap. A simple mean-gain approximation gives the projected forcing spectrum
entering the reduced SDE (Appendix~\ref{app:reduced}),
\begin{equation}
S_{\xi}(\omega)
\approx
\frac{\sigma_n^2}{N}\,
g^2\,\chi_{2,x}\,
|\hat{G}_{\mathrm{mf}}(i\omega)|^2\,
S_\phi^*(\omega).
\label{eq:Sxi}
\end{equation}
The observed coherent-mode spectrum is further shaped by the coherent feedback
loop,
\begin{equation}
S_{\kappa}^{\mathrm{obs}}(\omega)
\approx
\frac{S_{\xi}(\omega)}
     {|1-L(i\omega)|^2},
\end{equation}
away from the Hopf boundary.

We represent the projected colored forcing by a driven damped oscillator
$\xi(t)$ with natural frequency $\omega_\xi$ and damping ratio
$\zeta_{\mathrm{d}}$ matched to the peak and half-width of $S_{\xi}$, and
white-noise amplitude
$\sigma_\xi^2 = 4\zeta_{\mathrm{d}}\omega_\xi^3
\int S_{\xi}(\omega)\,\dd\omega/(2\pi)$, chosen so that the variance of
$\xi$ matches the projected forcing variance. The full stochastic reduced
system is
\begin{equation}
\begin{aligned}
\dd\kappa &= \frac{-\kappa + G_{\mathrm{eff}}(\lambda_{\mathrm{eff}}\kappa - \kappa_a) + \xi}{\tau_m}\,\dd t,\\
\dd\kappa_a &= \frac{-\kappa_a + \beta\kappa}{\tau_a}\,\dd t,\\
\dd Q &= \frac{-Q + Q_{\mathrm{fp}}(\Sigma_h^2,\beta)}{\tau_c}\,\dd t,\\
\dd\xi &= v\,\dd t,\\
\dd v &= (-2\zeta_{\mathrm{d}}\omega_\xi v - \omega_\xi^2\xi)\,\dd t + \sigma_\xi\,\dd W_t.
\end{aligned}
\label{eq:rd_SDE}
\end{equation}
Below the chaos threshold $S_\phi^* = 0$ and the stochastic terms vanish.

The noise amplitude depends on the local saturation: at the fixed point
$\bar{\chi}_x = 1 - Q$ (since $\sech^2(x) + \tanh^2(x) = 1$), so the noise
power (through $\chi_{2,x}$) and the deterministic stability (through
$\bar{\chi}_{\mathrm{eff}}$) are both governed by the degree of saturation.
When most neurons are saturated, the noise is weak and
oscillations are regular. As $\beta$ increases and the trajectory explores
lower-saturation regions, the noise grows and regularity is lost, accounting
for the transition from regular to irregular oscillations.

Figure~\ref{fig:phase_3d_panel} compares the three-dimensional
$(\kappa,\kappa_a,\Sigma_h^2)$ trajectories from full network simulations (top
row) with those of the stochastic reduced model~\eqref{eq:rd_SDE} (bottom
row). The reduced model evaluates all coefficients at the DMFT fixed point, so
the effective Hopf boundary is shifted relative to the full network. The
$\beta$ values for the middle column are chosen to match the qualitative
regime rather than to correspond numerically. At $\beta = 0.45$ (left column), the network is above the chaos threshold but
the coherent mode is a stable focus. Both the RNN and the SDE show the
trajectory confined to one basin, spiraling around the fixed point. The chaotic
bulk provides the energy that sustains this oscillation. At intermediate
$\beta$ (middle column), the stability margin has shrunk and chaotic
fluctuations occasionally drive the trajectory across the separatrix, producing
irregular switching between the two symmetric basins. Just beyond the bifurcation, the orbit winds
tightly around each (now unstable) fixed point before departing to the opposite
well, producing a characteristic lobed structure. As $\beta$ increases further,
this local winding unravels and the limit cycle develops the profile of a
relaxation oscillation, with slow drift along the $\kappa_a$ nullcline
punctuated by fast transitions between the two wells (right column), reflecting the timescale
separation $\tau_a \gg \tau_m$.

\section{Extensions}
\label{sec:extensions}
The theory extends to rank-$R$ connectivity with $K$ parallel adaptation
channels. We outline the main results. The derivations follow the same steps as
the single-channel case with straightforward algebraic modifications.

\subsection{Multiple adaptation channels}
We first describe how the fixed-point and spectral theory change when each
neuron carries several adaptation variables.
Each neuron carries $K$ adaptation variables with distinct timescales and
strengths:
\begin{equation}
\tau_k\,\dot{a}_i^{(k)} = -a_i^{(k)} + \beta_k\,\phi(x_i), \qquad k = 1,\dots,K,
\end{equation}
with $\beta_{\mathrm{tot}} = \sum_k\beta_k$. At steady state
$\sum_k a_i^{(k)*} = \beta_{\mathrm{tot}}\phi(x_i^*)$, so the fixed-point
diffeomorphism $F(x) = x + \beta_{\mathrm{tot}}\tanh(x)$ and the quantities
$\bar{\chi}_{\mathrm{eff}}$, $G_{\mathrm{eff}}$, and $Q_{\mathrm{fp}}$ depend
only on $\beta_{\mathrm{tot}}$, not on how adaptation is distributed across
channels. The fixed-point statistics are therefore identical to those of a
single channel with $\beta = \beta_{\mathrm{tot}}$.

The dynamics, however, are sensitive to the distribution of timescales and
strengths. Eliminating the $K$ adaptation variables from the linearized
equations gives the rational single-neuron denominator
\begin{equation}
D_{\mathrm{rat}}^{(K)}(s;c)
=
s\tau_m + 1
+
c\sum_{k=1}^K \frac{\beta_k}{s\tau_k+1}.
\label{eq:DK_rat}
\end{equation}
Equivalently, multiplying by $\prod_k(s\tau_k+1)$ gives the polynomial
denominator
\begin{equation}
D_{\mathrm{poly}}^{(K)}(s;c)
=
(s\tau_m + 1)\prod_{k=1}^K(s\tau_k + 1)
+
c\sum_{k=1}^K\beta_k\prod_{\ell\neq k}(s\tau_\ell + 1).
\label{eq:DK}
\end{equation}
The transfer function is therefore
\begin{equation}
\hat{G}^{(K)}(s;c)
=
\frac{1}{D_{\mathrm{rat}}^{(K)}(s;c)}
=
\frac{\prod_k(s\tau_k+1)}
     {D_{\mathrm{poly}}^{(K)}(s;c)}.
\label{eq:GK}
\end{equation}
The adaptation channels therefore do not act as independent output filters.
They combine into a single scalar negative-feedback kernel. For
$\beta_k\geq 0$, the imaginary part of
$D_{\mathrm{rat}}^{(K)}(i\omega;c)$ satisfies
\begin{equation}
\frac{1}{\omega}\im D_{\mathrm{rat}}^{(K)}(i\omega;c)
=
\tau_m
-
c\sum_{k=1}^K
\frac{\beta_k\tau_k}{1+\omega^2\tau_k^2},
\end{equation}
which is monotone increasing in $\omega^2$. Thus the feedback phase can cancel
the membrane phase at most once. Multiple adaptation channels can shift,
broaden, or skew the dominant resonant band, but in this shared-drive,
same-sign adaptation model they do not produce one independent spectral peak per channel.
\subsection{Rank-$R$ connectivity}
We then extend the coherent-mode description from a scalar overlap to a vector
of overlaps.
The loading vectors are drawn as
$(m_i^1,\dots,m_i^R,n_i^1,\dots,n_i^R)\overset{\mathrm{iid}}{\sim}
\mathcal{N}(0,\Sigma)$ where $\Sigma$ is a $2R\times 2R$ positive-definite
covariance with cross-covariance block $\Sigma_{nm}$. For each mode $\mu = 1,\dots,R$, define the overlap
$\kappa_\mu = N^{-1}\sum_j n_j^\mu\phi(x_j)$ and collect them into the vector
$\bm{\kappa} = (\kappa_1,\dots,\kappa_R)^\top$. Stein's lemma applied to each
component gives the self-consistency
$\bm{\kappa} = \bar{\chi}_{\mathrm{eff}}\,\Sigma_{nm}\bm{\kappa}$. Since $\bar{\chi}_{\mathrm{eff}}$ is real, only real eigenvalues of
$\Sigma_{nm}$ can satisfy this condition. Complex eigenvalue pairs do not produce stationary overlaps but can drive time-dependent oscillations of $\bm{\kappa}$ even without any adaptation
terms~\cite{MO18}. This is a distinct mechanism from the one
studied here. In higher-rank networks with a rotational structured component,
oscillations are built into the connectivity geometry, and their frequency is
set by the complex outlier eigenvalues of the structured connectivity. In the
rank-one model studied above, the structured connectivity supplies only a real
coherent mode. The phase lag needed for oscillation is instead generated by
single-neuron adaptation, so the relevant timescale is tied to $\tau_a$.

\subsection{Spectral and coherent boundaries}
This subsection gives the bulk and coherent stability conditions for the
general rank-$R$, $K$-adaptation-channel model.
The bulk spectral boundary is independent of the rank but depends on the
single-neuron transfer function. In terms of the rational denominator,
\begin{equation}
g^2\,\E\!\left[
\frac{c^2}{|D_{\mathrm{rat}}^{(K)}(s;c)|^2}
\right] = 1.
\end{equation}
Equivalently, using the polynomial denominator in~\eqref{eq:DK},
\begin{equation}
g^2\,\E\!\left[
c^2
\frac{\left|\prod_{k=1}^K(s\tau_k+1)\right|^2}
     {|D_{\mathrm{poly}}^{(K)}(s;c)|^2}
\right] = 1.
\end{equation}
In the mean-field approximation, parameterizing
$\mu = g\sqrt{\chi_{2,x}}\,e^{i\theta}$:
\begin{multline}
(s\tau_m + 1)\!\prod_{k=1}^K(s\tau_k+1)
  + \bar{\chi}_x\!\sum_{k=1}^K\beta_k\!\prod_{l\neq k}(s\tau_l+1) \\
= \mu\!\prod_{k=1}^K(s\tau_k+1).
\label{eq:si_parametric}
\end{multline}
Defining
\begin{equation}
K_{\mu\nu}(s)
=
\E\!\left[
n^\mu m^\nu\,
c_{\mathrm{lin}}(h)\,
\hat{G}^{(K)}\!\big(s;c_{\mathrm{lin}}(h)\big)
\right],
\end{equation}
the coherent eigenvalues are determined by
\begin{equation}
\det[I-K(s)] = 0.
\end{equation}
Only in special cases, for example when the loading covariance is diagonalized
and the operating-point distribution decouples the modes, does this reduce to
independent scalar equations
$\lambda_\alpha\E[c_{\mathrm{lin}}\hat{G}^{(K)}(s;c_{\mathrm{lin}})]=1$.

\subsection{Low-dimensional dynamics}
Finally, we state the corresponding reduced order-parameter dynamics for the general model.
The Gaussian closure of Sec.~\ref{sec:reduced} generalizes to $R$ overlaps
each carrying $K$ adaptation channels, plus the incoherent variance $Q$, giving
a system of dimension $R(K+1)+1$. For each mode $\mu = 1,\dots,R$, define the
adaptation overlaps
$\kappa_{a,\mu}^{(k)} = N^{-1}\sum_j n_j^\mu a_j^{(k)}$. The reduced
equations are
\begin{equation}
\begin{aligned}
\tau_m\,\dot{\kappa}_\mu &= -\kappa_\mu + G_{\mathrm{eff}}\!\left(
  \sum_\nu[\Sigma_{nm}]_{\mu\nu}\kappa_\nu
  - \sum_k\kappa_{a,\mu}^{(k)}\right),\\
\tau_k\,\dot{\kappa}_{a,\mu}^{(k)} &= -\kappa_{a,\mu}^{(k)}
  + \beta_k\,\kappa_\mu,\\
\tau_c\,\dot{Q} &= -Q + Q_{\mathrm{fp}}(\Sigma_h^2,\beta_{\mathrm{tot}}),
\end{aligned}
\label{eq:si_rd}
\end{equation}
with $\Sigma_h^2 = \bm{\kappa}^\top\Sigma_{mm}\bm{\kappa} + g^2Q$. The first
equation replaces $\lambda_{\mathrm{eff}}\kappa - \kappa_a$ in the rank-one
case~\eqref{eq:rd_3d} by its natural multi-mode generalization. Linearizing
around a fixed point along eigenvector $\alpha$ of $\Sigma_{nm}$ and
eliminating the $K$ adaptation perturbations gives the characteristic
polynomial
\begin{equation}
(s\tau_m + 1 - \lambda_\alpha G_{\mathrm{eff}})\!\prod_{k=1}^K(s\tau_k+1)
  + G_{\mathrm{eff}}\!\sum_{k=1}^K\beta_k\!\prod_{l\neq k}(s\tau_l+1) = 0,
\label{eq:si_charpoly}
\end{equation}
which reduces to~\eqref{eq:rd_chareq} at $R = K = 1$. Above chaos, the projected forcing cross-spectrum generalizes
Eq.~\eqref{eq:Sxi} to
\begin{equation}
S_{\xi_\mu\xi_\nu}(\omega) \approx
  \frac{[\Sigma_{nn}]_{\mu\nu}}{N}\,g^2\chi_{2,x}\,
  |\hat{G}_{\mathrm{mf}}^{(K)}(i\omega)|^2\,S_\phi^*(\omega).
\label{eq:si_cross}
\end{equation}
The $1/N$ scaling of this finite-size forcing persists for every pair
$(\mu,\nu)$.

\section{Discussion}
\label{sec:discussion}

We have shown that adding a single slow adaptation current to a low-rank
random network is sufficient to produce four qualitatively distinct dynamical
regimes: a static coherent state, noise-sustained oscillations that progress
from regular to irregular, stochastic switching between symmetric wells, and a
global limit cycle. The adaptation equation
$\tau_a\dot{a} = -a + \beta\phi(x)$ introduces one parameter ($\beta$) and one
timescale ($\tau_a$) per neuron, yet the resulting phase diagram is
comparable in richness to that of Brunel's excitatory-inhibitory
network~\cite{Brunel2000}, which requires two cell types with distinct synaptic
kinetics.

The dynamical mean-field theory reveals the mechanism behind this progression. Adaptation introduces a phase lag in the
single-neuron transfer function $\hat{G}(i\omega;c)$ that converts the
coherent mode from a stable node into a damped oscillator with resonant
frequency set by $\tau_a$. The overlap noise spectrum
$S_\kappa \propto |\hat{G}_{\mathrm{mf}}|^2 S_\phi^*/N$ shows quantitatively
how the coherent mode filters the chaotic bulk fluctuations. The transfer
function $\hat{G}$ selectively amplifies noise near the adaptation resonance. The two instability boundaries
organize the phase diagram. When chaos occurs first, the network passes through
noise-sustained oscillation and switching before reaching a global limit cycle;
when the Hopf bifurcation occurs first, the network jumps directly to global
oscillation.

Throughout all four regimes, individual neurons fire at heterogeneous rates set
by their projection onto the connectivity pattern. Even in the global limit
cycle, where the population overlap $\kappa(t)$ oscillates coherently, each
neuron's firing rate is modulated around a different operating point $x_i^*$,
and above the chaos threshold these single-neuron trajectories are further
corrupted by network-generated noise from the random connectivity. The
coexistence of population-wide temporal structure with single-neuron
irregularity and heterogeneity present in cortical
recordings~\cite{Wang2010}.

The DMFT for random networks~\cite{SCS88,CrisantiSompolinsky2018} and its
extension to low-rank
structure~\cite{MO18,Kadmon2015,Schuessler2020,LandauSompolinsky2018,LandauSompolinsky2021}
provide a well-developed framework for analyzing chaos and low-dimensional
dynamics in recurrent circuits. Within this framework, coherent oscillations
have required either rank-two or higher connectivity with appropriate
geometry~\cite{MO18} or complex outlier
eigenvalues~\cite{LandauSompolinsky2018}. Linear membrane-potential-driven adaptation in purely random networks
produces a resonant spectral peak but no spatial
organization~\cite{MGS19}. The firing-rate-driven form adopted here,
motivated by the spike-triggered potassium currents underlying
SFA~\cite{BendaHerz2003}, interacts with the random bulk and low-rank structure to produce
rich coherent oscillatory dynamics. The
temporal structure is set by the adaptation timescale rather than by the
connectivity geometry, and the full sequence of dynamical regimes unfolds by
varying a single parameter.

At large adaptation strength, the global limit cycle carries the network
between two symmetric states with slow transitions mediated by the adaptation
variable, resembling the Up--Down state alternations observed during slow-wave
sleep and anesthesia~\cite{Compte2003,Destexhe1993}. At intermediate
adaptation, the noise-sustained oscillations of the coherent mode produce
waxing-and-waning rhythmic episodes reminiscent of sleep
spindles~\cite{Destexhe1993}. A possible physiological
interpretation is that wakefulness, with stronger cholinergic suppression of
adaptation currents, corresponds to smaller effective $\beta$, whereas sleep
and anesthesia can involve larger effective adaptation through reduced
cholinergic drive~\cite{mccormick1993actions, stiefel2008cholinergic, nghiem2020cholinergic}. Note thta this mapping is speculative. Brain state transitions also involve changes in synaptic drive, thalamocortical coupling, inhibition, and neuromodulatory systems not included in the model.

The model thus captures, within a single
architecture and a one-dimensional parameter sweep, oscillatory phenomenology
that spans multiple physiological states. Whether the quantitative features of
these rhythms can be matched to
experimental recordings from specific brain states is an open question that
would benefit from fitting the model to neural data. More broadly, the
theoretical framework developed here may inform the design of recurrent neural
network architectures that produce robust and controllable oscillations, by
identifying adaptation strength and timescale as tuning knobs that are
complementary to the connectivity structure.

\begin{acknowledgments}
We thank Adam J. Eisen, Matthew B. Broschard, and Alicia KY. Lu for helpful discussions and comments on the manuscript.
This work was supported by Army Research Office W911NF2410228, Freedom Together Foundation, and The Picower Institute for Learning and Memory. We acknowledge the MIT Office of Research Computing and Data (ORCD) for providing high-performance computing resources that have contributed to the research results reported in this paper.
\end{acknowledgments}

\section*{Data Availability Statement}
The code that supports the findings of this article is openly available at
\url{https://github.com/Bowen-Zheng-99/rnn_adapt}.

\appendix
\allowdisplaybreaks
\section{Derivation of the parametric spectral boundary}
\label{app:spectral}

We detail the derivation of the parametric spectral boundary and the chaos thresholds stated in Sec.~\ref{sec:spectral}.

Within the mean-field approximation~\eqref{eq:rho2_mf}, the spectral boundary $\rho^2(s) = 1$ reads
\begin{equation}
g^2\chi_{2,x} = |D_{\mathrm{mf}}(s)|^2.
\label{eq:app_rho1}
\end{equation}
Writing $D_{\mathrm{mf}}(s) = s + 1 + \beta\bar{\chi}_x/(s\tau_a + 1)$ and multiplying through by $(s\tau_a + 1)$:
\begin{equation}
(s + 1)(s\tau_a + 1) + \beta\bar{\chi}_x = \mu(s\tau_a + 1),
\label{eq:app_rho2}
\end{equation}
where $\mu = g\sqrt{\chi_{2,x}}\,e^{i\theta}$ parameterizes the circle $|\mu| = g\sqrt{\chi_{2,x}}$. Expanding the left-hand side:
\begin{equation}
\tau_a s^2 + (1 + \tau_a)s + 1 + \beta\bar{\chi}_x = \mu\tau_a s + \mu.
\end{equation}
Rearranging yields the quadratic~\eqref{eq:parametric}. For each $\theta\in[0,2\pi)$, the two roots $s_\pm(\theta)$ trace a closed curve in the complex $s$-plane.

The chaos threshold $g_c$ corresponds to the rightmost crossing of the imaginary axis. Setting $s = i\omega$ in~\eqref{eq:app_rho1} and using $|D_{\mathrm{mf}}(i\omega)|^2 = 1/|\hat{G}(i\omega)|^2$:
\begin{equation}
g^2\chi_{2,x}\,|\hat{G}(i\omega)|^2 = 1.
\label{eq:app_gc_cond}
\end{equation}
The maximum over $\omega$ determines the minimum $g$ at which the boundary touches the imaginary axis, giving~\eqref{eq:gc}.

To find $\max_\omega|\hat{G}(i\omega)|^2$, write $u = \omega^2$ and define the numerator and denominator of~\eqref{eq:Ghat_sq} as
\begin{align}
R(u) &= 1 + \tau_a^2 u, \\
P(u) &= \tau_a^2 u^2 + (1 + \tau_a^2 - 2b\tau_a)\,u + (1+b)^2,
\end{align}
where $b \equiv \beta\bar{\chi}_x$, so that $|\hat{G}(i\omega)|^2 = R(u)/P(u)$.  Differentiating and setting $R'P - RP' = 0$:
\begin{equation}
\tau_a^4 u^2 + 2\tau_a^2 u - \big[\tau_a^2 b(2{+}b) + 2b\tau_a - 1\big] = 0.
\label{eq:app_u_quad}
\end{equation}
The positive root is
\begin{equation}
u_* = \frac{-1 + \sqrt{\tau_a^2 b(2{+}b) + 2b\tau_a}}{\tau_a^2}.
\end{equation}
An interior resonant maximum requires
$\tau_a^2 b(2{+}b) + 2b\tau_a > 1$.
For $\tau_a \gg 1$ with $b = O(1)$, this simplifies to
\begin{equation}
u_* \approx \frac{\sqrt{b(2+b)}}{\tau_a}, \qquad \omega_* = \sqrt{u_*} \approx \frac{[b(2+b)]^{1/4}}{\sqrt{\tau_a}}.
\end{equation}

To evaluate the peak height, note that at the critical point $R'P = RP'$ implies $|\hat{G}(i\omega_*)|^2 = R/P = \tau_a^2/P'(u_*)$, where $P'(u) = 2\tau_a^2 u + (1 + \tau_a^2 - 2b\tau_a)$.  Substituting $u_* = \sqrt{b(2+b)}/\tau_a + O(1/\tau_a^2)$:
\begin{equation}
P'(u_*) = \tau_a^2 + 2\tau_a\sqrt{b(2+b)} - 2b\tau_a + O(1),
\end{equation}
so that
\begin{equation}
|\hat{G}(i\omega_*)|^2 = \frac{\tau_a}{\tau_a + \Delta(b)} + O(1/\tau_a^2),
\end{equation}
where
\begin{equation}
\Delta(b) = 2\big[\sqrt{b(2+b)} - b\big].
\end{equation}
Since $\sqrt{b(2+b)} - b = 2b/(\sqrt{b(2+b)}+b) \to 1$ as $b \to \infty$, we have $0 \leq \Delta \leq 2$ for all $b \geq 0$.  Inserting into~\eqref{eq:app_gc_cond} yields the oscillatory threshold~\eqref{eq:gosc}.

For $\omega = 0$, $|\hat{G}(0)|^2 = 1/(1+b)^2$, giving the static threshold~\eqref{eq:gstat}. Since $g_{\mathrm{stat}} = (1+b)/\sqrt{\chi_{2,x}}$ grows linearly with $b$ while $g_{\mathrm{osc}}$ is bounded above by $\sqrt{(\tau_a+2)/\tau_a}/\sqrt{\chi_{2,x}}$, the oscillatory threshold is lower for sufficiently large $b$.

\section{Hermite expansion for the spectral self-consistency}
\label{app:hermite}

We derive the spectral self-consistency map $\mathcal{T}: S_\phi \mapsto S_\phi'$ used in Sec.~\ref{sec:spectrum}.

Write $\delta x_i(t) \equiv x_i(t) - x_i^*$ and $\delta a_i(t) \equiv a_i(t)
- a_i^*$ for the perturbations around the fixed point. Subtracting the
fixed-point equation from~\eqref{eq:dmft_x}, the constant rank-one input
$m_i\kappa^*$ drops out. In the $N\to\infty$ limit the overlap self-averages
to $\kappa^*$, so the remaining coherent-mode contribution
$m_i(\kappa(t)-\kappa^*)$ vanishes and the perturbation is driven solely by
$g\,\delta\eta_i(t) \equiv g(\eta_i(t) - \eta_i^{(0)})$, the zero-mean
component of the Gaussian input, whose power spectral density is $S_\phi(f)$.
Linearizing with $c_i = \sech^2(x_i^*)$, the perturbations satisfy
\begin{align}
\tau_m\,\delta\dot{x}_i &= -\delta x_i - \delta a_i + g\,\delta\eta_i(t), \label{eq:app_lin_x}\\
\tau_a\,\delta\dot{a}_i &= -\delta a_i + \beta c_i\,\delta x_i. \label{eq:app_lin_a}
\end{align}
Taking Fourier transforms and eliminating $\delta\hat{a}_i$ gives
\begin{equation}
D_i(i\omega)\,\delta\hat{x}_i(\omega) = g\,\delta\hat{\eta}_i(\omega),
\end{equation}
where $D_i(s) = s\tau_m + 1 + \beta c_i/(s\tau_a + 1)$ is the single-neuron
characteristic function~\eqref{eq:Di}. The membrane-potential fluctuation
spectrum and variance are therefore
\begin{align}
S_{\delta x,i}(f) &= g^2\,|\hat{G}_i(i\omega)|^2\,S_\phi(f), \label{eq:app_Sdx}\\
\sigma_{x,i}^2 &= 2g^2\int_0^\infty |\hat{G}_i(i\omega)|^2\,S_\phi(f)\,\dd f, \label{eq:app_var}
\end{align}
with $\hat{G}_i(i\omega) = 1/D_i(i\omega)$ and normalized autocorrelation
$r_i(\tau) = C_{\delta x,i}(\tau)/\sigma_{x,i}^2$. 

Since $\delta x_i$ is a zero-mean Gaussian process, and hence a linear filter of the Gaussian input $\delta\eta_i$, we can write $\delta x_i(t) = \sigma_{x,i}\,z_i(t)$ where $z_i$ is a unit-variance stationary Gaussian process with autocorrelation $r_i(\tau)$. The firing rate is then a deterministic nonlinear function of a Gaussian argument, $\phi_i(t) = \tanh(x_i^* + \sigma_{x,i}\,z_i(t))$, which we expand in the probabilist Hermite polynomials $\He_p$ ($\He_0 = 1$, $\He_1 = z$, $\He_2 = z^2 - 1$, etc.):
\begin{equation}
\tanh(x_i^* + \sigma_{x,i}z) = \sum_{p=0}^\infty \frac{b_p^{(i)}}{p!}\,\He_p(z),
\label{eq:app_hermite}
\end{equation}
with coefficients
\begin{equation}
b_p^{(i)} = \E_z\!\big[\tanh(x_i^* + \sigma_{x,i}z)\,\He_p(z)\big].
\end{equation}
The $p=0$ term is the time-averaged firing rate, $b_0^{(i)} = \E_z[\tanh(x_i^* + \sigma_{x,i}z)] = \langle\phi_i\rangle_t$, and the centered fluctuation is $\delta\phi_i = \sum_{p\geq 1} b_p^{(i)}\He_p(z_i)/p!$. The Mehler formula expands the joint density of $(z_1, z_2)$ with zero means,
unit variances, and correlation $\rho$ as
\begin{equation}
p(z_1, z_2) = \varphi(z_1)\,\varphi(z_2)
  \sum_{p=0}^\infty \frac{\rho^p}{p!}\,\He_p(z_1)\,\He_p(z_2),
\label{eq:app_mehler}
\end{equation}
where $\varphi$ is the standard Gaussian density. Multiplying both sides by
$\He_p(z_1)\,\He_q(z_2)$, integrating, and using
$\int \He_p\,\He_q\,\varphi\,dz = p!\,\delta_{pq}$ gives
\begin{equation}
\E\!\big[\He_p(z(t))\,\He_q(z(t+\tau))\big] = \delta_{pq}\,p!\,r(\tau)^p
\label{eq:app_isserlis}
\end{equation}
for any stationary unit-variance Gaussian process with autocorrelation
$r(\tau)$. This identity underlies the self-consistent autocorrelation
equations in random network
DMFT~\cite{SCS88,CrisantiSompolinsky2018}.

Substituting the Hermite expansion of $\delta\phi_i$ and
applying~\eqref{eq:app_isserlis}, the cross terms ($p \neq q$) vanish and the
per-neuron firing-rate autocovariance becomes
\begin{equation}
C_{\phi,i}(\tau) = \sum_{p=1}^P \frac{(b_p^{(i)})^2}{p!}\,r_i(\tau)^p,
\label{eq:app_Cphi_i}
\end{equation}
where $P$ is the truncation order. The $p = 1$ term, proportional to
$r_i(\tau)$, is the linearized contribution; the $p = 2$ term, proportional to
$r_i(\tau)^2$, generates spectral energy at twice the peak frequency, and so
on.

All neuron-specific quantities ($c_i$, $\sigma_{x,i}$, $r_i$, $b_p^{(i)}$)
are deterministic functions of the quenched input $h_i$. The fixed point
$x^* = F^{-1}(h)$ determines the local gain
$c = \sech^2(x^*)$~\eqref{eq:local_gain_def}, which enters the transfer
function and sets $\sigma_x$ and $r$
through~\eqref{eq:app_Sdx}--\eqref{eq:app_var}, and $b_p$ follows
from~\eqref{eq:app_hermite}. Since $h\sim\mathcal{N}(0,\Sigma_h^2)$, the
population-averaged autocovariance is
\begin{equation}
C_\phi(\tau) = \E_h\!\left[\sum_{p=1}^P
  \frac{(b_p(h))^2}{p!}\,r(h,\tau)^p\right],
\label{eq:app_Cphi_pop}
\end{equation}
a one-dimensional Gaussian integral. Its cosine transform yields a new spectrum
$S_\phi'(f)$, completing the spectral map
$\mathcal{T}: S_\phi \mapsto S_\phi'$.

We now verify that the spectral map reproduces the chaos threshold from
Sec.~\ref{sec:spectral}. Near onset, $\sigma_{x,i} \to 0$ for all neurons.
Expanding $\tanh(x_i^* + \sigma_{x,i}z)$ in powers of $\sigma_{x,i}$ gives
$b_1^{(i)} = \sigma_{x,i}\,c_i + O(\sigma_{x,i}^3)$ and
$b_p^{(i)} = O(\sigma_{x,i}^p)$ for $p \geq 2$. Since the $p$-th term
in~\eqref{eq:app_Cphi_i} scales as $b_p^2 \sim \sigma_{x,i}^{2p}$, only
$p = 1$ survives. At this order the Hermite expansion reduces to linearization,
$\delta\phi_i \approx c_i\,\delta x_i$, so the per-neuron firing-rate spectrum
is $S_{\delta\phi,i}(f) = c_i^2\,g^2\,|\hat{G}_i(i\omega)|^2\,S_\phi(f)$.
Population-averaging,
\begin{equation}
S_\phi'(f) = g^2\,\E\!\big[c^2|\hat{G}(i\omega;c)|^2\big]\,S_\phi(f)
  = \rho^2(i\omega)\,S_\phi(f),
\end{equation}
where $\rho^2(i\omega)$ is the spectral radius from~\eqref{eq:rho2}. The
linearized spectral map thus multiplies $S_\phi$ frequency-by-frequency by
$\rho^2$. When $\rho^2 < 1$ at all frequencies, the trivial fixed point
$S_\phi = 0$ is stable. A nontrivial solution $S_\phi^* \neq 0$ bifurcates
when $\max_\omega\rho^2(i\omega) = 1$. This is the same condition derived from
the eigenvalue analysis in Sec.~\ref{sec:spectral}, confirming that the two
approaches give the same chaos threshold.

\section{Fluctuation corrections to the overlap self-consistency}
\label{app:fluct_corrections}

Above chaos ($g > g_c$), each neuron fluctuates around its fixed point $x_i^* = F^{-1}(h_i)$ with amplitude $\sigma_x(h_i) > 0$ determined by the spectral DMFT (Sec.~\ref{sec:spectrum}). The overlap must therefore be computed from the time-averaged firing rates $\langle\phi_i\rangle_t$ rather than the static values $\phi(x_i^*)$. Under the DMFT, $x_i(t) = x_i^* + \delta x_i(t)$ where $\delta x_i$ is a zero-mean Gaussian process with variance $\sigma_{x,i}^2$, so
\begin{equation}
\langle\phi_i\rangle_t = \E_\zeta\!\big[\tanh\!\big(x_i^* + \sigma_{x,i}\,\zeta\big)\big], \qquad \zeta\sim\mathcal{N}(0,1),
\label{eq:app_phi_avg}
\end{equation}
which is the $p = 0$ Hermite coefficient $b_0^{(i)}$ from Appendix~\ref{app:hermite}. The overlap $\kappa = N^{-1}\sum_j n_j\langle\phi_j\rangle_t$ involves jointly Gaussian $(n_j, h_j)$ with $\Cov(n,h) = \lambda_{\mathrm{eff}}\kappa$, and $\langle\phi_j\rangle_t$ depends on $h_j$ alone. Stein's lemma gives
\begin{equation}
\kappa = \lambda_{\mathrm{eff}}\kappa\,\E_h\!\big[\partial_h\langle\phi\rangle_t\big].
\label{eq:app_stein_fluct}
\end{equation}
For $\kappa\neq 0$, this yields $1/\lambda_{\mathrm{eff}} = \E_h[\partial_h\langle\phi\rangle_t]$.

The time-averaged rate $\langle\phi\rangle_t$ depends on $h$ through two channels. The fixed point $x^* = F^{-1}(h)$ and the fluctuation amplitude $\sigma_x(h)$. The chain rule gives
\begin{align}
\partial_h\langle\phi\rangle_t &= \E_\zeta\!\Big[\sech^2\!\big(x^* + \sigma_x\zeta\big)\,\Big((F^{-1})'(h) + \sigma_x'(h)\,\zeta\Big)\Big] \nonumber\\
&= (F^{-1})'(h)\,\tilde{c}(h) + \sigma_x'(h)\,\E_\zeta\!\big[\sech^2(x^* + \sigma_x\zeta)\,\zeta\big],
\label{eq:app_dhphi}
\end{align}
where $\tilde{c}(h) = \E_\zeta[\sech^2(x^* + \sigma_x\zeta)]$ is the time-averaged gain~\eqref{eq:eff_gain}. The first term, using $(F^{-1})'(h) = 1/(1 + \beta c(h))$ from~\eqref{eq:Finv_deriv}, equals $\tilde{c}(h)/(1 + \beta\,c(h))$. The numerator carries $\tilde{c}$ because shifting $x^*$ shifts the entire fluctuating trajectory; the denominator carries the static $c(h)$ because $(F^{-1})'$ is a property of the deterministic equation $F(x) = h$.

For the second term, Gaussian integration by parts ($\E[\zeta\,g(\zeta)] = \E[g'(\zeta)]$) gives
\begin{multline}
\E_\zeta\!\big[\sech^2(x^* + \sigma_x\zeta)\,\zeta\big] \\
= -2\sigma_x\,\E_\zeta\!\big[\sech^2(x^* + \sigma_x\zeta)\tanh(x^* + \sigma_x\zeta)\big].
\label{eq:app_hermite_coeff}
\end{multline}
This is $O(\sigma_x)$. Meanwhile $\sigma_x'(h) = O(\sigma_x)$ because $\sigma_x^2(h) \propto S_\phi^*$, which vanishes at the chaos threshold, so $\partial_h\sigma_x^2 = O(\sigma_x^2)$ near onset and $\sigma_x' = (\partial_h\sigma_x^2)/(2\sigma_x) = O(\sigma_x)$. The second term is therefore $O(\sigma_x^2)$, and we drop it.

Retaining only the first term and averaging over the population,
\begin{equation}
\frac{1}{\lambda_{\mathrm{eff}}} \approx \E_h\!\left[\frac{\tilde{c}(h)}{1 + \beta\,c(h)}\right],
\label{eq:app_eff_sc}
\end{equation}
which is Eq.~\eqref{eq:eff_selfcons}. The mean-squared firing rate is similarly updated to
\begin{equation}
Q = \E_h\!\big[\E_\zeta\!\big[\tanh^2\!\big(F^{-1}(h) + \sigma_x(h)\zeta\big)\big]\big].
\end{equation}
Below chaos ($\sigma_x = 0$), $\tilde{c}(h) = c(h)$ and~\eqref{eq:app_eff_sc} reduces to $\lambda_{\mathrm{eff}}\bar{\chi}_{\mathrm{eff}} = 1$, recovering~\eqref{eq:selfcons_fp}.

\section{Coherent eigenvalue and self-consistent gain}
\label{app:coherent}

We derive the coherent loop gain~\eqref{eq:loop_tf} that determines the rank-one outlier eigenvalue.

Returning to the full characteristic equation~\eqref{eq:Ndim_chareq} with
$J_{ij} = m_i n_j/N + g W_{ij}/\sqrt{N}$, the perturbation
$\delta x_i \propto e^{st}$ satisfies
\begin{equation}
D_i(s)\,\delta x_i = \frac{m_i}{N}\sum_j n_j c_j\,\delta x_j
  + \frac{g}{\sqrt{N}}\sum_j W_{ij} c_j\,\delta x_j.
\label{eq:app_full_eig}
\end{equation}
Define $\delta\kappa_s = N^{-1}\sum_j n_j c_j \delta x_j$. Rearranging:
\begin{equation}
\delta x_i = \frac{m_i\,\delta\kappa_s}{D_i(s)}
  + \frac{g}{D_i(s)\sqrt{N}}\sum_j W_{ij} c_j\,\delta x_j.
\label{eq:app_dx_recur}
\end{equation}
For $s$ outside the bulk eigenvalue region ($\rho^2(s) < 1$), the resolvent
$R(s) = [I - g\,\diag(D_i(s))^{-1}\,WC/\sqrt{N}]^{-1}$ is well-defined.
Solving~\eqref{eq:app_dx_recur} formally:
$\delta x_i = \delta\kappa_s \sum_k R_{ik}(s)\,m_k/D_k(s)$.
Projecting onto $\bm{n}$ and dividing by $\delta\kappa_s \neq 0$:
\begin{equation}
1 = \frac{1}{N}\sum_{i,k} \frac{n_i c_i\,R_{ik}(s)\,m_k}{D_k(s)}.
\label{eq:app_proj}
\end{equation}
Expanding $R$ in a Neumann series, the $p=0$ (diagonal) term gives
$N^{-1}\sum_i n_i m_i c_i/D_i(s)$. For $p \geq 1$, each factor of
$W_{ab}/\sqrt{N}$ contributes $O(N^{-1/2})$. Since $n_i$ and $m_k$ are drawn
independently of $W$, the expectation of
$n_i W_{ij_1}\cdots W_{j_{p-1}k}m_k$ with all distinct indices vanishes; the
leading nonvanishing contributions require index coincidences, which are
$O(N^{-1})$ per pair. The off-diagonal terms are therefore $O(N^{-p/2})$ and
vanish as $N\to\infty$, giving
\begin{equation}
1 = \E\!\left[\frac{nm\,c}{D(s;c)}\right],
\label{eq:app_diag_lim_static}
\end{equation}
which is~\eqref{eq:coherent_limit}. Above chaos, we approximate the static
gain $c(h)$ by the fluctuation-averaged gain $\tilde{c}(h)$
from~\eqref{eq:eff_gain}, replacing $c\,\hat{G}(s;c)$ by
$\tilde{c}\,\hat{G}(s;\tilde{c})$. This treats each neuron as if it were
operating at a modified fixed point with gain $\tilde{c}$ rather than $c$. The
approximation is exact below chaos where $\tilde{c} = c$, and introduces errors
of $O(\sigma_x^2)$ above chaos:
\begin{equation}
1 \approx \E\!\big[nm\,\tilde{c}\,\hat{G}(s;\tilde{c})\big].
\label{eq:app_diag_lim}
\end{equation}

\subsection{Gaussian conditioning}

The expectation in~\eqref{eq:app_diag_lim} is over the joint distribution of
$(n,m,h)$. These are jointly Gaussian with zero mean and covariances
$\Cov(n,m) = \gamma\sigma_m\sigma_n = \lambda_{\mathrm{eff}}$,
$\Cov(n,h) = \lambda_{\mathrm{eff}}\kappa$,
$\Cov(m,h) = \sigma_m^2\kappa$.
Since $\tilde{c}$ and $\hat{G}$ depend on $h$ alone, we condition on $h$ using
the Gaussian regression formula:
\begin{align}
n \mid h &= \frac{\lambda_{\mathrm{eff}}\kappa}{\Sigma_h^2}\,h + \tilde{n},
  \label{eq:app_n_cond}\\
m \mid h &= \frac{\sigma_m^2\kappa}{\Sigma_h^2}\,h + \tilde{m},
  \label{eq:app_m_cond}
\end{align}
where $\tilde{n}$ and $\tilde{m}$ are residuals independent of $h$. The
conditional expectation of $nm$ is
\begin{align}
\E[nm \mid h]
  &= \frac{\lambda_{\mathrm{eff}}\kappa}{\Sigma_h^2}\,
     \frac{\sigma_m^2\kappa}{\Sigma_h^2}\,h^2
     + \E[\tilde{n}\tilde{m}] \nonumber\\
  &= \lambda_{\mathrm{eff}}\alpha\,w^2
     + \lambda_{\mathrm{eff}}(1-\alpha),
\label{eq:app_Enm}
\end{align}
where the residual covariance is
$\E[\tilde{n}\tilde{m}] = \lambda_{\mathrm{eff}}
  - \lambda_{\mathrm{eff}}\sigma_m^2\kappa^2/\Sigma_h^2
  = \lambda_{\mathrm{eff}}(1-\alpha)$,
with $\alpha = \sigma_m^2\kappa^2/\Sigma_h^2$ and $w = h/\Sigma_h$.
Inserting into~\eqref{eq:app_diag_lim} via the law of iterated expectations
gives the coherent loop gain~\eqref{eq:loop_tf}:
\begin{equation}
L(s) =
\lambda_{\mathrm{eff}}\,
\E_w\!\left[
\big((1{-}\alpha)+\alpha w^2\big)\,
c_{\mathrm{lin}}(\Sigma_h w)\,
\hat{G}\!\big(s;c_{\mathrm{lin}}(\Sigma_h w)\big)
\right].
\end{equation}
The corresponding outlier condition is $L(s)=1$.

At $s = 0$, $\hat{G}(0;c)=1/(1+\beta c)$, so
\begin{equation}
L(0) =
\lambda_{\mathrm{eff}}\,
\E\!\left[
  \frac{(1-\alpha+\alpha w^2)c_{\mathrm{lin}}}
       {1+\beta c_{\mathrm{lin}}}
\right].
\end{equation}
Below chaos, $c_{\mathrm{lin}}=c$. Defining
\begin{equation}
f(w) =
\frac{c(\Sigma_h w)}
     {1+\beta c(\Sigma_h w)},
\end{equation}
the zero-frequency loop gain becomes
\begin{equation}
L(0)
=
\lambda_{\mathrm{eff}}
\left[
\E[f(w)] + \alpha\,\E[(w^2-1)f(w)]
\right].
\end{equation}
On the nonzero fixed-point branch, Eq.~\eqref{eq:selfcons_fp} gives
$\lambda_{\mathrm{eff}}\E[f(w)]=1$. Therefore
\begin{equation}
L(0)-1
=
\lambda_{\mathrm{eff}}\alpha\,
\E[(w^2-1)f(w)].
\end{equation}
This correction is generally nonzero because the gain depends on the
operating point. 

\section{Low-dimensional dynamics closure}
\label{app:reduced}

We derive the reduced equations~\eqref{eq:rd_3d} and the projected forcing
spectrum~\eqref{eq:Sxi}. The construction follows the Gaussian closure approach of the low-rank RNN literature~\cite{MO18,Schuessler2020}.

Define the projected membrane potential
$\kappa_x(t) \equiv N^{-1}\sum_i n_i x_i(t)$. This is distinct from the
overlap $\kappa = N^{-1}\sum_j n_j\phi(x_j)$; relating the two requires a
closure. Multiplying~\eqref{eq:model_x} by $n_i/N$, summing over $i$, and
using $N^{-1}\sum_i n_i m_i = \lambda_{\mathrm{eff}} + O(N^{-1/2})$:
\begin{equation}
\tau_m\,\dot{\kappa}_x = -\kappa_x + \lambda_{\mathrm{eff}}\kappa - \kappa_a + \text{(noise)},
\label{eq:app_kappax_raw}
\end{equation}
where the noise term $N^{-1}\sum_i n_i g\eta_i$ is the projection of the
random input onto $\bm{n}$, treated below.

Under the instantaneous Gaussian ansatz, the joint distribution of
$(n_i, x_i)$ across the population at each time is approximated by its DMFT
fixed-point form. In particular, $\phi(x_i) \approx \psi(h_{\mathrm{eff},i})$
where $h_{\mathrm{eff},i} = m_i\kappa - a_i$ and
$\psi(h) = \tanh(F^{-1}(h))$. Since
$\Cov(n, h_{\mathrm{eff}}) = \lambda_{\mathrm{eff}}\kappa - \kappa_a$,
Stein's lemma gives
\begin{equation}
\kappa = N^{-1}\sum_j n_j\,\phi(x_j) \approx (\lambda_{\mathrm{eff}}\kappa - \kappa_a)\,\bar{\chi}_{\mathrm{eff}}.
\label{eq:app_stein_dyn}
\end{equation}
This motivates the relaxation equation
$\tau_m\dot{\kappa} = -\kappa + G_{\mathrm{eff}}(\lambda_{\mathrm{eff}}\kappa - \kappa_a)$.
Using $G_{\mathrm{eff}} = \bar{\chi}_{\mathrm{eff}}$ directly, the fixed-point
condition ($\dot{\kappa} = \dot{\kappa}_a = 0$, $\kappa_a = \beta\kappa$)
gives $(\lambda_{\mathrm{eff}} - \beta)\bar{\chi}_{\mathrm{eff}} = 1$, which
differs from the DMFT condition
$\lambda_{\mathrm{eff}}\bar{\chi}_{\mathrm{eff}} = 1$. The error arises
because the closure treats $\kappa_a$ as an additive external parameter,
whereas in the DMFT the adaptation enters through $F^{-1}$ and couples
nonlinearly to the gain. Requiring the relaxation equation to reproduce the
DMFT fixed point yields
\begin{equation}
G_{\mathrm{eff}} = \frac{1}{\lambda_{\mathrm{eff}} - \beta}
  = \frac{\bar{\chi}_{\mathrm{eff}}}{1 - \beta\bar{\chi}_{\mathrm{eff}}},
\label{eq:app_Geff_derive}
\end{equation}
which is Eq.~\eqref{eq:Geff}. At $\beta = 0$,
$G_{\mathrm{eff}} = \bar{\chi}_{\mathrm{eff}}$ and the standard MO18 closure
is recovered.

The incoherent variance $Q = E[\phi^2]$ depends
on $\Sigma_h^2 = \sigma_m^2\kappa^2 + g^2Q$ through
\[
Q_{\mathrm{fp}}(\Sigma_h^2,\beta)
= E_Z\!\left[\tanh^2\!\left(F^{-1}(\Sigma_h Z)\right)\right].
\]
A perturbation $\delta Q$ induces $\delta \Sigma_h^2 = g^2 \delta Q$,
so linearizing the self-consistency map gives
\[
\delta Q' \approx g^2 Q'_{\mathrm{fp}}(\Sigma_h^{2*},\beta)\,\delta Q.
\]
Thus contractivity requires $g^2 Q'_{\mathrm{fp}} < 1$.
Since $Q$ involves $\phi^2$, its bare decay rate is approximated as
$2/\tau_{\mathrm{auto}}$ (the $p=2$ Hermite contribution decays as
$r(\tau)^2$; see Appendix~\ref{app:hermite}), where
\[
\tau_{\mathrm{auto}} = \frac{S_\phi^*(0)}{2C_\phi(0)}.
\]
Matching the linearized relaxation rate gives
\[
\tau_c \dot Q = -Q + Q_{\mathrm{fp}}(\Sigma_h^2,\beta), \qquad
\tau_c \approx
\frac{\tau_{\mathrm{auto}}}
{2\left(1 - g^2 Q'_{\mathrm{fp}}(\Sigma_h^{2*},\beta)\right)}.
\]
Above chaos, the incoherent fluctuations provide a forcing of
the coherent overlap. Linearizing
$\delta\phi_i \approx c_i\,\delta x_i$ gives
$\delta\kappa(t) = N^{-1}\sum_i n_i\,c_i\,\delta x_i(t)$.
Since each $\delta x_i$ is driven by an independent Gaussian process $\eta_i$,
cross-neuron correlations vanish at leading order and the projected forcing
spectrum is
\begin{align}
S_{\xi}(\omega)
  &= \frac{1}{N^2}\sum_i n_i^2 c_i^2\,S_{\delta x,i}(\omega) \nonumber\\
  &= \frac{1}{N}\,
     \E[n^2 c^2\,g^2|\hat{G}(i\omega;c)|^2]\,
     S_\phi^*(\omega) \nonumber\\
  &\approx
     \frac{\sigma_n^2}{N}\,
     g^2\,\E[c^2|\hat{G}(i\omega;c)|^2]\,
     S_\phi^*(\omega).
\label{eq:app_Sxi_full}
\end{align}
The last step replaces $\E[n^2\mid h]$ by $\sigma_n^2$, dropping the
conditioning correction from the structured branch. Replacing
$\E[c^2|\hat{G}|^2]$ by
$\chi_{2,x}\,|\hat{G}_{\mathrm{mf}}|^2$ gives the simplified expression in
Eq.~\eqref{eq:Sxi}.
\renewcommand{\footnoterule}{}
\bibliography{references}

\end{document}